\documentclass[sn-basic]{sn-jnl}

\usepackage{bm}
\usepackage{fix-cm}
\usepackage{xfrac}
\usepackage{mathrsfs}
\usepackage{bold-extra}
\DeclareFontFamily{U}{rsfs}{\skewchar\font127}
\DeclareFontShape{U}{rsfs}{m}{n}{
   <-6> rsfs5
   <6-8> rsfs7
   <8-> rsfs10
}{}

\usepackage{graphicx}
\usepackage{multirow}
\usepackage{amsmath,amssymb,amsfonts}
\usepackage{amsthm}
\usepackage{mathrsfs}
\usepackage[title]{appendix}
\usepackage{xcolor}
\usepackage{textcomp}
\usepackage{manyfoot}
\usepackage{booktabs}
\usepackage{algorithm}
\usepackage{algorithmicx}
\usepackage{algpseudocode}
\usepackage{listings}

\newcommand{\araa}{Annu. Rev. Astron. Astrophys.} 
\newcommand{\aj}{Astron. J.} 
\newcommand{\apj}{Astrophys. J.} 
\newcommand{\apjs}{Astrophys. J. Suppl. Ser.} 
\newcommand{\apss}{Astrophys. Space Sci.} 
\newcommand{\aap}{Astron. Astrophys.} 
\newcommand{\nat}{Nature} 
\newcommand{\pasp}{Publ. Astron. Soc. Pac.}

\graphicspath{{./}{figures/}}

\begin{document}

\title[H$\alpha$ variability of solar-like stars]{Variability of H$\alpha$ chromospheric activity of solar-like stars revealed by the time-domain data of LAMOST Medium-Resolution Spectroscopic Survey}

\author*[1,2]{\fnm{Han} \sur{He}}\email{hehan@nao.cas.cn}

\author[1,2,3]{\fnm{Ali} \sur{Luo}}

\author[1,2,3]{\fnm{Haotong} \sur{Zhang}}

\author[1,2,3]{\fnm{Song} \sur{Wang}}

\affil[1]{\orgname{National Astronomical Observatories, Chinese Academy of Sciences}, \orgaddress{\city{Beijing}, \postcode{100101}, \country{China}}}

\affil[2]{\orgname{University of Chinese Academy of Sciences}, \orgaddress{\city{Beijing}, \postcode{100049}, \country{China}}}

\affil[3]{\orgname{CAS Key Laboratory of Optical Astronomy, Chinese Academy of Sciences}, \orgaddress{\city{Beijing}, \postcode{100101}, \country{China}}}

\abstract{
The variability of H$\alpha$ chromospheric activity of solar-like stars is investigated by using the time-domain data of LAMOST Medium-Resolution Spectroscopic Survey (MRS).
Strict screening conditions are employed to ensure the quality of the selected MRS spectra and the consistency between the H$\alpha$ spectra of each stellar source.
We use $R_\mathrm{H\alpha}$ index (ratio of H$\alpha$ luminosity to bolometric luminosity) to measure the H$\alpha$ activity intensity of a spectrum,
and utilize the median of the $R_\mathrm{H\alpha}$ values of multiple observations ($R_\mathrm{H\alpha}^\mathrm{median}$) as the representative activity intensity of a stellar source.
The H$\alpha$ variability of a stellar source is indicated by the extent of $R_\mathrm{H\alpha}$ fluctuation ($R_\mathrm{H\alpha}^\mathrm{EXT}$) of multiple observations.
Our sample shows that $R_\mathrm{H\alpha}^\mathrm{EXT}$ of solar-like stars is about one order of magnitude smaller than $R_\mathrm{H\alpha}^\mathrm{median}$.
The distribution of $\log R_\mathrm{H\alpha}^\mathrm{EXT}$ versus $\log R_\mathrm{H\alpha}^\mathrm{median}$ reveals the distinct behaviors between the stellar source categories with lower ($\log R_\mathrm{H\alpha}^\mathrm{median} < -4.85$) and higher ($\log R_\mathrm{H\alpha}^\mathrm{median} > -4.85$) activity intensity.
For the former stellar source category, the top envelope of the distribution first increases and then decreases with $\log R_\mathrm{H\alpha}^\mathrm{median}$;
while for the latter category, the top envelope of the distribution is largely along a positive correlation line.
In addition, for the stellar sources with lower activity intensity, 
the large-$\log R_\mathrm{H\alpha}^\mathrm{EXT}$ objects near the top envelope of the $\log R_\mathrm{H\alpha}^\mathrm{EXT}$ versus $\log R_\mathrm{H\alpha}^\mathrm{median}$ distribution tend to have long-term and regular variations of H$\alpha$ activity;
while for the stellar sources with higher activity intensity, 
the H$\alpha$ variations are more likely to be random fluctuations.
}

\keywords{
Stellar activity,
Stellar chromospheres,
Solar-like star
}

\maketitle

\section{Introduction} \label{sec:intro}

Stellar magnetic activity is believed to originate from the interaction between stellar rotation and internal convection,
and has been detected in the stellar convective envelope as well as in the convective core through the asteroseismology approach (e.g., \citealt{2009ARA&A..47..333D, 2016Natur.529..364S}). 
The strengths of the magnetic fields of different stars can span several orders of magnitude, 
which are influenced by the stellar types, abundances, rotation rates and evolution stages (e.g., \citealt{2009ARA&A..47..333D}).
The stars in their very early evolution stages are believed to have stronger magnetic activity owing to the faster rotation,
and the magnetic fields in turn can influence the evolution of stars (e.g. \citealt{1972ApJ...171..565S, 2015Natur.517..589M}).
Intense magnetic activities lead to enhanced ultraviolet and X-ray emissions from stellar atmospheres,
which can influence the habitability of planets around stars as well as the chemical components of planetary atmospheres (e.g., \citealt{2020ApJ...902..114W, 2024ApJS..273....8H, 2025ApJS..276...29L}).

The magnetic field activity of stars can be manifested through various features in the stellar atmosphere,
e.g., the dark starspots in the stellar photosphere and the bright plages in the stellar chromosphere.
The latter can be detected via the radiation at the center of the chromospheric spectral lines (e.g., \citealt{2017ARA&A..55..159L}),
such as the Ca II H and K lines, Ca II infrared triplet lines, and hydrogen H$\alpha$ line (e.g., \citealt{2022ApJS..263...12Z, 2024A&A...688A..23Z, 2024ApJS..272....6H, 2023Ap&SS.368...63H, 2024NatSR..1417962H}).
Higher line-core fluxes of the chromospheric lines indicate the stronger magnetic field activity of stars,
thus the line-core fluxes of the chromospheric lines are often employed as the proxies of stellar magnetic activity (e.g., \citealt{2008LRSP....5....2H}).

One of the most prominent observation projects on stellar chromospheric activity is the long-term monitoring of the stellar Ca II H and K lines at the Mount Wilson Observatory (MWO) (\citealt{1978ApJ...226..379W}).
The decades of observation data of MWO, with the observing cadence as short as one day, 
revealed the long-term cyclic variations of stellar activity (i.e., the stellar activity cycles) (e.g., \citealt{1995ApJ...438..269B, 1998ASPC..154..153B}) as well as the short-term variations of stellar activity caused by the stellar rotational modulation (e.g., \citealt{1981ApJ...250..276V, 1983ApJ...275..752B}).
The existence of rotational modulation indicates that brightness in the stellar chromosphere is not homogeneous (due to the patchy magnetic field distribution on the stellar surface), 
whereas the long-term variation of stellar activity associated with stellar activity cycles is more essential (related to the stellar dynamo mechanisms). 
In the time-series diagrams of the stellar Ca II H and K signals published in the literature
(e.g., \citealt{1983ApJ...275..752B, 1995ApJ...438..269B, 2022AJ....163..183B}),
the fluctuations due to the rotational modulation have timescales related to the stellar rotation period, while the long-term variations of stellar activity due to the stellar activity cycles have timescales of years to decades.

In addition to the rotational modulation and stellar activity cycles, 
another research interest based on the long-term observation data is the variability of stellar activity, that is,
the dispersion of activity fluctuations of individual stars.
The concerns are on the relationship between the stellar activity intensity and the activity variability. 
The stellar activity intensity can be described by the mean or median of the time series of a certain activity index (e.g., the widely used $R'_\mathrm{HK}$ index for the Ca II H and K lines),
while the activity variability can be measured by the standard deviation (STD; i.e., rms of the differences from the mean) or other dispersion metric of the activity index time series.
In the works by \citet{2007ApJS..171..260L} and \citet{2009AJ....138..312H},
a trend of positive correlation between $\log [\mathrm{mean}(R'_\mathrm{HK})]$ and $\log [\mathrm{STD}(R'_\mathrm{HK})]$ is shown for the solar-like stars.

Like the Ca II H and K lines, the hydrogen H$\alpha$ line is another commonly used spectral line to indicate chromospheric activity of solar-like stars (e.g., \citealt{1979ApJ...234..579C, 1985ApJ...289..269H, 1991A&A...251..199P, 2005A&A...431..329L, 2009A&A...501.1103M}).
For F-type and G-type stars,
the H$\alpha$ line mainly exhibits an absorption-line profile, 
and the enhanced line-core radiation due to stellar activity manifests as the filling-in of the absorption line.
For K-type and later type stars,
if the activity intensity is high enough,
the line-core radiation may exceed the continuum and the H$\alpha$ line can exhibit an emission-line profile (e.g., \citealt{1990ApJS...74..891R}).
\citet{2014A&A...566A..66G, 2022A&A...668A.174G} and \citet{2022A&A...658A..57M} analyzed the variability of H$\alpha$ activity of FGK main-sequence stars using the long-term (several years) observation data of the HARPS spectrograph and investigated the correlation between the stellar H$\alpha$ activity and Ca II H and K activity.

In 2018, the LAMOST telescope \citep{2012RAA....12.1197C, 2012RAA....12.1243L} initiated a dedicated time-domain spectroscopic survey project.
This time-domain survey project is part of the Medium-Resolution Spectroscopic Survey (MRS) of LAMOST,
with a spectral resolution power ($\lambda / \Delta\lambda$) of about 7500 \citep{2020arXiv200507210L}.
Each MRS spectrum has two observational bands, the red band (about 6300--6800\,\AA) and the blue band (about 4950--5350\,\AA), 
and the H$\alpha$ line (vacuum wavelength: 6564.614\,\AA; \citealt{2002AJ....123..485S}) is well covered by the red band of MRS.
Thus the time-domain spectral data of the H$\alpha$ line are available in the recent data releases of LAMOST, 
which provides an opportunity to study stellar H$\alpha$ variability \citep{2021RNAAS...5....6H}.
In this work, we investigate the variability of H$\alpha$ chromospheric activity of solar-like stars based on the time-domain data of MRS.

The content of this paper is organized as follows.
In Section \ref{sec:data}, we explain the time-domain data of LAMOST-MRS.
In Section \ref{sec:sample}, we describe the detailed procedure of sample selection for the MRS time-domain spectra and sources of solar-like stars employed in this work.
In Section \ref{sec:measure}, we describe the processing of MRS H$\alpha$ line data and the measures employed in this work to quantitatively indicate the stellar H$\alpha$ activity intensity and variability.
In Section \ref{sec:result}, we present the analysis results of the variability of H$\alpha$ chromospheric activity of solar-like stars based on the MRS time-domain data.
In Section \ref{sec:discu}, we give further discussions on the analysis results.
Section \ref{sec:con} is the conclusion of this work.

\section{Time-domain data of LAMOST-MRS} \label{sec:data}

The LAMOST-MRS is designed to consist of two parts of surveys, the non time-domain (NT) survey and the time-domain (TD) survey.
The NT survey emphasizes the uniform distribution of the observed objects in the sky, 
and the TD survey repeatedly observes the same groups of objects in pre-selected fields for many times over the years \citep{2020arXiv200507210L}.
It should be noted that, 
for NT survey, it is possible that some objects are observed multiple times by coincidence through different observation proposals,
while for TD survey, both the cadence and the number of observations are not uniform.

Each MRS observation for a stellar object contains multiple single-exposures.
According to the observation plan of MRS, 
the duration of each single-exposure is about 1200 seconds;
NT survey takes 3 single-exposures for each observation, and TD survey takes 3--8 single-exposures \citep{2020arXiv200507210L}. 
In practical observations, more or less single-exposures are also possible.
These single-exposure spectra are then coadded to obtain the spectral data with higher single-to-noise ratio. 
Both the single-exposure spectra and the coadded spectrum of one MRS observation are stored in the same FITS data file and share the same {\tt\string obsid} (unique LAMOST observation identifier).
Each single-exposure spectrum also has its own observation identifier (named {\tt\string mobsid} in the LAMOST catalogs).
The single-exposure spectral data can be used to analyze transient events of stars, such as stellar flares (e.g., \citealt{2022ApJ...928..180W}).
In this study, we are concerned with the stellar H$\alpha$ variability of steady chromosphere instead of transient events,
thus the coadded spectral data of MRS are utilized in this work for their higher single-to-noise ratio than the single-exposure spectra.
The observation time of a coadded spectrum is defined as the median time of multiple single-exposures, which is provided by the keyword {\tt\string DATE-OBS} in the MRS FITS file.

The MRS data used in this work are from the LAMOST Data Release 10 (DR10)\footnote{\url{https://www.lamost.org/dr10/v1.0/}}, 
which encompasses the data from the first five years of MRS (up to June 2022). 
We identify different stellar sources through {\tt uid} (unique LAMOST source identifier) provided in the LAMOST data release catalogs.
We do not intentionally separate the NT- and TD-survey data. 
If a stellar object was observed three or more times by MRS (having three or more coadded spectra in the MRS catalogs),
it is considered as a time-domain object and the associated MRS spectra are considered as the time-domain spectra.
The selection procedure for the MRS time-domain sample of solar-like stars employed in this work and the statistical properties of the selected MRS time-domain sample will be described in detail in Section~\ref{sec:sample}.

\begin{figure}
  \centering
  \includegraphics[width=1.16\textwidth]{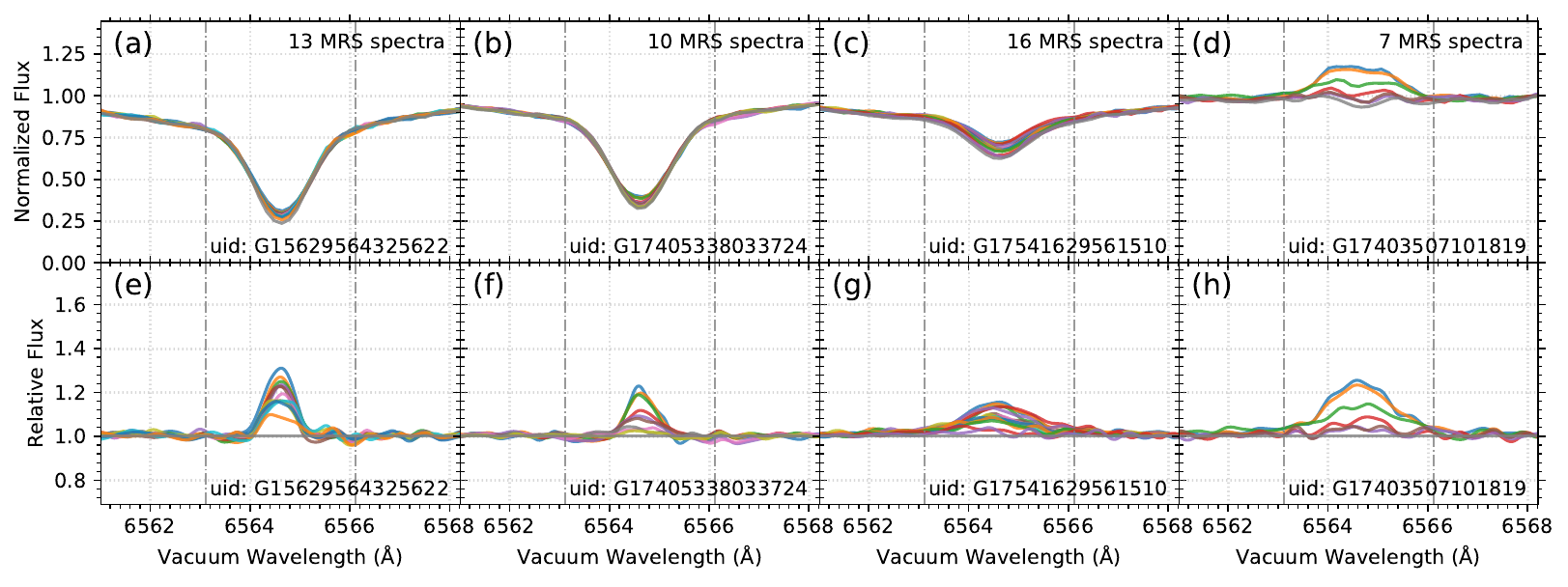}
  \caption{Example of the MRS time-domain spectra of H$\alpha$ line of solar-like stars.
    The four columns show the MRS time-domain H$\alpha$ spectra of four example stellar sources, respectively,
  with the activity levels of the stellar sources increasing from left to right.
  The {\tt\string uid} and the number of MRS spectra of each stellar source are printed. 
  The different spectra of the same stellar source are displayed in a variety of colors for ease of distinction.
  In each column, the top row shows the MRS H$\alpha$ spectra normalized by the continuum, 
  and the bottom row shows the relative flux of the H$\alpha$ spectra with respect to the reference spectrum (the spectrum with the lowest H$\alpha$ central flux; displayed in gray).
  The wavelength shift caused by radial velocity is corrected.
  The time spans of the time-domain spectra of the four stellar sources are all greater than 1,000 days.
  The two dash-dot lines in each panel enclose a 3\,{\AA}-wide wavelength range around the H$\alpha$ line center, 
  within which the variations of the H$\alpha$ profiles caused by stellar activity mainly happen.
  }
  \label{fig:mrs_timedomain_spectra_example}
\end{figure}

Figure \ref{fig:mrs_timedomain_spectra_example} gives an example of the MRS time-domain spectra of H$\alpha$ line of solar-like stars. 
A total of four example stellar sources are employed in Figure \ref{fig:mrs_timedomain_spectra_example}.
Each column of Figure~\ref{fig:mrs_timedomain_spectra_example} is for one stellar source,
with the {\tt\string uid} and the number of MRS spectra of the stellar source being printed; 
the top row shows the MRS H$\alpha$ spectra normalized by the continuum, 
and the bottom row shows the relative flux of the H$\alpha$ spectra with respect to the reference spectrum (the spectrum with the lowest H$\alpha$ central flux; displayed in gray).
The activity levels of the stellar sources shown in Figure~\ref{fig:mrs_timedomain_spectra_example} increase from left to right.
The time spans of the time-domain spectra of the four stellar sources are all greater than 1,000 days.
It can be seen from Figure~\ref{fig:mrs_timedomain_spectra_example} that the variations of the H$\alpha$ profiles caused by stellar activity mainly happen within the wavelength range of about 3\,{\AA} around the line center (enclosed by two dash-dot lines in each panel of Figure \ref{fig:mrs_timedomain_spectra_example}).
In line wings, the spectral profiles of the same stellar source are almost identical.

\section{Sample selection} \label{sec:sample}

The {\tt LAMOST MRS Parameter Catalog} of LAMOST DR10 provides stellar atmospheric parameters (effective temperature $T_\mathrm{eff}$, surface gravity $\log g$, and metallicity $\mathrm{[Fe/H]}$) determined by the LAMOST stellar parameter pipeline (LASP; \citealt{2015RAA....15.1095L}) as well as other stellar and spectroscopic parameters (such as radial velocity, projected rotational velocity, etc.) for more than two millions of MRS coadded spectra.
The MRS time-domain sample of solar-like stars employed in this work is selected from this catalog.
Strict screening conditions are employed to ensure the quality of the selected MRS spectra and the consistency between the H$\alpha$ spectra of each stellar source.
The screening conditions for MRS spectra are described in Section \ref{sec:sample_spcond},
and the screening conditions for stellar sources are described in Section \ref{sec:sample_srccond}.
In Section \ref{sec:sample_statistics}, we give statistical properties of the selected MRS time-domain sample of solar-like stars.

\subsection{Screening conditions for MRS spectra} \label{sec:sample_spcond}

\subsubsection{Solar-like condition} \label{sec:sample_spcond_solarlike}
The solar-like sample is selected by the stellar atmospheric parameters of the MRS spectra provided in the {\tt LAMOST MRS Parameter Catalog}.
In this work, we adopt the solar-like condition of $4500\,\mathrm{K} < T_\mathrm{eff} < 6500\,\mathrm{K}$ and $\log g > 4.1\,\mathrm{dex}$, 
which includes the late F-type, G-type, and early K-type main-sequence stars.

Since the stellar atmospheric parameters $T_\mathrm{eff}$, $\log g$, and $\mathrm{[Fe/H]}$ ({\tt teff\_lasp}, {\tt logg\_lasp}, and {\tt feh\_lasp} parameters in the catalog) and their uncertainties (denoted by $\delta T_\mathrm{eff}$, $\delta \log g$, and $\delta \mathrm{[Fe/H]}$) are extensively used in this work,
we only keep the MRS spectra with all three atmospheric parameters and their uncertainties being available in the catalog.
The spectra with one or more missing atmospheric parameters or their uncertainties are removed from our sample.

\subsubsection{Signal-to-noise ratio condition} \label{sec:sample_spcond_snr}
The MRS spectra with high signal-to-noise ratio (S/N) are required to reduce the uncertainties of the derived activity indices of the H$\alpha$ line (see Section \ref{sec:measure}) as well as to reduce the uncertainties of the stellar atmospheric parameters.
In this work, for the red band of MRS (denoted by $R$; which contains the H$\alpha$ line) we use the S/N condition of $\mathrm{S/N}_R \ge 80$; 
if the stellar atmospheric parameters in the {\tt LAMOST MRS Parameter Catalog} are determined by the blue band data of MRS, the S/N condition for the blue band (denoted by $B$) is further requested, which is $\mathrm{S/N}_B \ge 50$.
This S/N condition is optimized for the analysis of MRS H$\alpha$ spectra of F-, G-, and K-type stars (see \citealt{2023Ap&SS.368...63H}).
Note that the {\tt LAMOST MRS Parameter Catalog} only provides the S/N values ({\tt snr} parameter in the catalog) for the MRS band ($R$ or $B$) utilized to determine the stellar atmospheric parameters.
The complete information of $\mathrm{S/N}_R$ and $\mathrm{S/N}_B$ can be found in the {\tt LAMOST MRS General Catalog}.

\subsubsection{Removing low-quality spectra} \label{sec:sample_spcond_quality}
Some MRS spectra have bad pixels in their spectral flux data, 
which can be identified by flux values equal to or less than zero.
Even if the flux data are valid, 
some pixels may also be marked as low quality or bad profile by the data quality columns ({\tt ANDMASK} and {\tt ORMASK} columns) in the MRS FITS data array.
A MRS coadded spectrum is generated from several single-exposures (see Section \ref{sec:data}).
If at lease one exposure has problem, the {\tt ORMASK} will be set;
if there is a problem with every exposure, the {\tt ANDMASK} will be set (see the LAMOST data release documentation for more details).
In this work, we use the {\tt ORMASK} column to identify low-quality data since it is more strict than the {\tt ANDMASK} column.
If a MRS spectrum has bad pixels (flux values equal to or less than zero) or low-quality pixels (indicated by {\tt ORMASK}) in the wavelength bands of the H$\alpha$ line utilized in this work, 
the spectrum is removed from our sample. 

The radial velocity parameter of a MRS spectrum is required to transform the wavelengths of the spectrum to the values in the rest frame. 
The {\tt LAMOST MRS Parameter Catalog} provides radial velocity parameters {\tt rv\_r0} and {\tt rv\_b0} for the red band and blue band of MRS, respectively.
Since the H$\alpha$ line is contained in the red band of MRS,
we use {\tt rv\_r0} and its uncertainty ({\tt rv\_r0\_err}) in the analysis of this work,
and the spectra with missing {\tt rv\_r0} or {\tt rv\_r0\_err} values are removed from our sample.
To ensure the reliability of the {\tt rv\_r0} data,
we cross-check the values of {\tt rv\_r0} from the red band and the values of {\tt rv\_b0} from the blue band.
For most MRS spectra, the values of {\tt rv\_r0} and {\tt rv\_b0} are consistent with each other.
A few MRS spectra do have large differences between their {\tt rv\_r0} and {\tt rv\_b0} values, which are considered to be unreliable.
In this work, we use the criterion of $|{\tt rv\_r0} - {\tt rv\_b0}| \ge 7$\,km/s to identify unreliable radial velocity data.
The threshold of the criterion (7\,km/s) corresponds to about one pixel of wavelength shift in the H$\alpha$ band of MRS. 
Those spectra with radial velocity data meeting the criterion are removed from our sample.

The moon light can also reduce the quality of the observed MRS spectra.
The {\tt LAMOST MRS Parameter Catalog} provides {\tt moon\_flg} and {\tt lunardate} parameters to help identify the spectra affected by the moon light. 
If the distance between an observed object and the moon is greater than 30 degree, 
the {\tt moon\_flg} parameter is set to 0,
otherwise it is set to 1.
The {\tt lunardate} parameter provides the observation date of a MRS spectrum in one lunar month according to the Chinese lunar calendar, 
which is convenient for determining the lunar phase associated with the spectrum.
The {\tt lunardate} of the full moon is about 15 (the mid date of a lunar month).
In this work, we identify the MRS spectra affected by the moon light by the condition of ${\tt moon\_flg}=1$ or
${\tt lunardate}=14$, 15, or 16.
The MRS spectra that meet this condition are removed from our sample.

\subsubsection{Removing spectra with large projected rotational velocities} \label{sec:sample_spcond_rotvel}
To minimize the influence of the rotational broadening effect on the H$\alpha$ line profile, 
the MRS spectra with large projected rotational velocity ($v \sin i$) should be avoided.
The {\tt LAMOST MRS Parameter Catalog} provides the $v \sin i$ values ({\tt vsini\_lasp} parameter in the catalog) for the MRS spectra with $v \sin i > 30$~km/s. 
For the spectra with $v \sin i$ less than 30~km/s, 
{\tt vsini\_lasp} parameter is not available and is set to $-9999.0$ in the catalog.
In this work, we only use the MRS spectra with $v \sin i$ less than 30~km/s, i.e., the spectra with {\tt vsini\_lasp} $=-9999.0$.
This condition has been quantitatively evaluated for the H$\alpha$ line data of MRS in \citet{2023Ap&SS.368...63H}.
The spectra that do not meet the condition are removed from our sample.

\subsubsection{Removing spectra contaminated by the H\,II regions} \label{sec:sample_spcond_HII} 
The H\,II regions are a type of Galactic nebulae ionized by the ultraviolet radiations from nearby young massive stars,
and have strong H$\alpha$ signals (e.g., \citealt{2017PASP..129h2001P, 2017PASP..129d3001P, 2021RAA....21...96W}).
If the observed stellar objects are close to or located in the H\,II regions,
the stellar H$\alpha$ fluxes may be contaminated with the signals from the H\,II regions.
Since the MRS includes the dedicated Galactic nebulae survey \citep{2021RAA....21...96W, 2022RAA....22g5015W},
the contamination of the stellar H$\alpha$ fluxes by the H\,II regions should be considered.

In this work, we identify the MRS spectra contaminated by the H\,II regions through the N\,II spectral lines on the two sides of the H$\alpha$ line. 
The N\,II lines are characteristic lines of H\,II regions (\citealt{2017PASP..129h2001P, 2017PASP..129d3001P}) and very weak for solar-like stars.
The vacuum wavelengths of the two N\,II lines are 6549.86\,{\AA} and 6585.27\,{\AA}, respectively \citep{2002AJ....123..485S},
which are covered by the red band of MRS (see example MRS spectrum of the N\,II lines from H\,II regions in Appendix \ref{sec:appendix_HII_regions}). 
We calculate the mean fluxes in the central bands of the N\,II lines,
and then normalize the central fluxes by the nearby continuum fluxes.
The results are named $N_V$ index and $N_R$ index for the two N\,II lines on the violet side and red side of the H$\alpha$ line, respectively.
By examining the relations of the $N_V$ and $N_R$ indices with the activity index of the H$\alpha$ line ($I_\mathrm{H\alpha}$ index; see definition in Section \ref{sec:measure_activityintensity}),
we find that the $N_R$ index is a better indicator to distinguish the spectra contaminated by the H\,II regions than the $N_V$ index.
The criterion adopted in this work for identifying the MRS spectra contaminated by the H\,II regions is $N_R \le 0.98$ or $N_R \ge 1.02$.
Details of the analysis to determine the criterion are provided in Appendix \ref{sec:appendix_HII_regions}.
The MRS spectra that meet the criterion are removed from our sample.

\subsection{Screening conditions for stellar sources} \label{sec:sample_srccond}

\subsubsection{Time-domain condition} \label{sec:sample_srccond_timedomain}
As explained in Section \ref{sec:data},
we use {\tt uid} provided in the LAMOST catalogs to distinguish between stellar sources observed by MRS.
Each record in the {\tt LAMOST MRS Parameter Catalog} corresponds to a MRS observation (and a coadded spectrum) and has a unique {\tt obsid}.
A stellar source ({\tt uid}) with three or more MRS observations ({\tt obsid}) after all the screening conditions described in Section \ref{sec:sample_spcond} is considered as a time-domain object employed in this work.
For clarity, we use $N_\mathrm{obs}$ to represent the number of MRS observations of a stellar source,
then the criterion for a time-domain object can be expressed as $N_\mathrm{obs} \ge 3$.
Only the stellar sources that meet this criterion are kept in our sample.

Another key parameter for a time-domain object is the time span of multiple observations (denoted by $T_\mathrm{span}$), 
which is defined as the time difference between the first and last MRS observations for the object. 
Since we are interested in stellar H$\alpha$ variability on time scales ranging from days to years,
the stellar sources with too short $T_\mathrm{span}$ scales are not preferred.
In this work, we only employ the stellar sources with time span of multiple observations greater than three days (i.e., $T_\mathrm{span} > 3$~days). 
The stellar sources that do not meet this condition are removed from our sample.

\subsubsection{Condition for consistency of stellar atmospheric parameters} \label{sec:sample_srccond_paraconsistency} 
In the {\tt LAMOST MRS Parameter Catalog}, the stellar atmospheric parameters $T_\mathrm{eff}$, $\log g$, and $\mathrm{[Fe/H]}$ are provided for each of the observations (coadded spectra) of the same stellar source.
Thus a time-domain object has multiple measurements of the stellar atmospheric parameters.
For the convenience of subsequent analysis, we assign a fixed group of stellar atmospheric parameters to each stellar source using the median of multiple measurements, 
that is, 
\begin{equation} \label{equ:def_source_paras}
  \left\{ \begin{array}{ll}
    T_\mathrm{eff,\,source} = \mathrm{median}(\{T_\mathrm{eff}^i\}) &  \\
    \log g_\mathrm{\,source} = \mathrm{median}(\{\log g^i\}) & \quad i=1, \cdots, N_\mathrm{obs}, \\
    \mathrm{[Fe/H]}_\mathrm{\,source} = \mathrm{median}(\{\mathrm{[Fe/H]}^i\}), &   
  \end{array} \right.
\end{equation}
where $\{T_\mathrm{eff}^i\}$, $\{\log g^i\}$, and $\{\mathrm{[Fe/H]}^i\}$ are the sets of the multiple measurements of the three stellar atmospheric parameters, respectively.
The uncertainties of the source parameters (denoted by $\delta T_\mathrm{eff,\,source}$, $\delta \log g_\mathrm{\,source}$, and $\delta \mathrm{[Fe/H]}_\mathrm{\,source}$) are inherited from the uncertainties of the median values.
The peaks of the $\delta T_\mathrm{eff,\,source}$, $\delta \log g_\mathrm{\,source}$, and $\delta \mathrm{[Fe/H]}_\mathrm{\,source}$ distributions for the stellar source sample after the above screening conditions are at about 26\,K, 0.024\,dex, and 0.016\,dex, respectively.

For most MRS time-domain objects, the deviations of the stellar parameter values of the multiple measurements from the source values defined in Equation (\ref{equ:def_source_paras}) are on the order of magnitude of the parameter uncertainties.
A few stellar sources do have large differences between the parameter values of multiple measurements,
which might be owing to accidentally wrong pointing of telescope fibers.
In order to recognize the objects with consistent stellar atmospheric parameters between multiple measurements, 
we define the deviation of one measurement, 
which is the absolute difference between the parameter value of the measurement and the source parameter value defined in Equation (\ref{equ:def_source_paras}).
For example, for the multiple measurements of effective temperature of a stellar source, 
$\{T_\mathrm{eff}^i, i=1, \cdots, N_\mathrm{obs}\}$,
the deviation of the $i$th measurement (denoted by $|\Delta T_\mathrm{eff}^i|$) is defined as $|\Delta T_\mathrm{eff}^i| = |T_\mathrm{eff}^i - T_\mathrm{eff,\,source} |$.
The deviations $|\Delta \log g^i|$ and $|\Delta \mathrm{[Fe/H]}^i|$ can be defined similarly.
The criterion for recognizing the stellar sources with consistent stellar atmospheric parameters is that the maximum deviation is less than five times uncertainty, that is,
\begin{equation} \label{equ:cond_para_deviation}
  \left\{ \begin{array}{ll}
    \max(\{|\Delta T_\mathrm{eff}^i|\}) < 5\times 26\,\mathrm{K} & \quad \textrm{and} \\
    \max(\{|\Delta \log g^i|\}) < 5\times 0.024\,\mathrm{dex} & \quad \textrm{and} \\
    \max(\{|\Delta \mathrm{[Fe/H]}^i|\}) < 5\times 0.016\,\mathrm{dex}. &
  \end{array} \right.
\end{equation}
The uncertainties values for effective temperature, surface gravity, and metallicity ($26$\,K, $0.024$\,dex, and $0.016$\,dex, respectively) used in Equation (\ref{equ:cond_para_deviation}) are the values corresponding to the peaks of the $\delta T_\mathrm{eff,\,source}$, $\delta \log g_\mathrm{\,source}$, and $\delta \mathrm{[Fe/H]}_\mathrm{\,source}$ distributions, as have been given above.
The stellar sources that do not meet this criterion are removed from our sample.

\subsubsection{Condition for consistency of H$\alpha$ profiles} \label{sec:sample_srccond_profileconsistency}

As illustrated in Figure \ref{fig:mrs_timedomain_spectra_example}, the variations of the H$\alpha$ profiles caused by stellar activity mainly happen in H$\alpha$ line center.
In line wings, the spectral profiles of multiple observations are almost identical.
For most MRS time-domain stellar objects, the H$\alpha$ profiles of multiple observations are consistent as the example shown in Figure~\ref{fig:mrs_timedomain_spectra_example}.
A few stellar sources do show discrepancies between the H$\alpha$ profiles of multiple observations,
which might be owing to the observation or calibration issues.
In the line wings, the discrepancy is mainly manifested as the offset of the normalized spectral fluxes (see example diagram in Appendix \ref{sec:appendix_profileconsistency_linewings}); 
and in the line center, the discrepancy is mainly manifested as the distortion of the spectral profiles (see example diagram in Appendix~\ref{sec:appendix_profileconsistency_linecenter}).

In order to recognize the objects with consistent H$\alpha$ profiles in the line wings, 
for each stellar source,
we first sort the H$\alpha$ spectra of multiple observations by their $I_\mathrm{H\alpha}$ index values (reflecting mean normalized flux at line center; see definition in Section \ref{sec:measure_activityintensity}),
and calculate the means of the normalized fluxes in two line-wing bands on the violet and red sides of the H$\alpha$ line for each of the spectra.
The central wavelengths of the two line-wing bands are $\pm 4.5$\,{\AA} from the H$\alpha$ line center,
and the widths of the two line-wing bands are both $2$\,{\AA} (see illustration in Appendix \ref{sec:appendix_profileconsistency_linewings}).
We then evaluate the differences of the wing-band mean fluxes between the H$\alpha$ spectrum with the smallest $I_\mathrm{H\alpha}$ index value and other spectra with larger $I_\mathrm{H\alpha}$ index values.
The criterion for recognizing the consistent H$\alpha$ profiles in the line wings is that the maximum absolute difference of the wing-band mean fluxes is less than 0.022 (see Appendix \ref{sec:appendix_profileconsistency_linewings} for details on determining the criterion).
The stellar sources that do not meet this criterion or have bad pixels in the line-wing bands of their H$\alpha$ spectra are removed from our sample. 

The situation of recognizing consistent H$\alpha$ profiles in the line center is more complicated than that in the line wings,
since the H$\alpha$ profiles in line center show different characteristics for stellar sources with different activity levels (see Figure~\ref{fig:mrs_timedomain_spectra_example}).
For the stellar sources with lower activity levels,
the H$\alpha$ profiles in line center are simple as demonstrated in Figures \ref{fig:mrs_timedomain_spectra_example}a and \ref{fig:mrs_timedomain_spectra_example}b;
while for the stellar sources with higher activity levels,
the H$\alpha$ profiles in line center are more complex as demonstrated in Figures \ref{fig:mrs_timedomain_spectra_example}c and \ref{fig:mrs_timedomain_spectra_example}d,
especially for the objects with H$\alpha$ central flux approaching continuum (see Figure \ref{fig:mrs_timedomain_spectra_example}d).
For this reason, we treat the stellar sources with lower and higher activity levels separately,
which are classified according to their $I_\mathrm{H\alpha}$ index values.
If the median of $I_\mathrm{H\alpha}$ of multiple observations (denoted by $I_\mathrm{H\alpha}^{\mathrm{median}}$) of a stellar source is less than $0.5$ (see Appendix \ref{sec:appendix_iha} and Figure \ref{fig:iha_vs_teff_mrs_timedomain_spectra}),
it is classified as having a lower activity level;
and a stellar source with $I_\mathrm{H\alpha}^{\mathrm{median}}$ greater than 0.5 is classified as having a higher activity level. 

In order to recognize the objects with consistent H$\alpha$ profiles in the line center from the class of stellar sources with lower activity levels,
for each stellar source in this class, we calculate the correlation coefficients of the normalized fluxes in a 3\,{\AA}-wide line-center band between the H$\alpha$ spectrum with the smallest $I_\mathrm{H\alpha}$ index value and other spectra with larger $I_\mathrm{H\alpha}$ index value.
The criterion for recognizing the consistent H$\alpha$ profiles in the line center is that the minimum correlation coefficient of the normalized fluxes in the center band is greater than 0.986 (see Appendix \ref{sec:appendix_profileconsistency_linecenter} for details on determining the criterion).
The stellar sources in the lower activity level class that do not meet this criterion or have bad pixels in the line-center band of their H$\alpha$ spectra are removed from our sample. 
For the stellar sources in the higher activity level class,
since the H$\alpha$ profiles in line center are more complex as explained above,
the consistency of the H$\alpha$ profiles in the line center is confirmed by visual inspection, 
and the criterion of the correlation coefficient in center band is not applied.

\subsubsection{Removing sources in the MRS catalog of close binaries} \label{sec:sample_srccond_binaries}

\citet{2024ApJ...976..243D} compared the stellar chromospheric activities of the binaries and the main sequence single stars by using the spectral data of the Ca II H and K lines of the LAMOST Low-Resolution Spectroscopic Survey.
Their results show that the binaries and single stars generally have a common distribution in the activity--rotation relation diagram, 
which suggests that the measured chromospheric activities of the binaries are from the more active components;
for the close binaries, there are clues that the strong interaction between the components of the binaries might affect the activity properties. 
For this reason, in this work we do not exclude binary stars in general, 
but only remove the sources of close binaries.

We utilize the catalog obtained in the work by \citet{2024ApJ...969..114L} to remove the sources of close binaries. 
The catalog is based on the same LAMOST-MRS data release (DR10 v1.0) as used in this work.
Although the work of \citet{2024ApJ...969..114L} is intended for the compact object candidates in binaries, 
the criterion adopted by the work (large or rapid radial-velocity variation) guarantees a reliable catalog from the perspective of close binaries.  
We cross-match our sample (after the above screening conditions) with the catalog of \citet{2024ApJ...969..114L} and find that six stellar sources are in common.
These common objects are removed from our sample.

\subsection{Statistical properties of the selected MRS time-domain sample of solar-like stars} \label{sec:sample_statistics}

There are 2,148,470 spectral records and 1,019,861 stellar sources in the original {\tt LAMOST MRS Parameter Catalog} used in this work.
After applying all the screening conditions for sample selection described in Sections \ref{sec:sample_spcond} and \ref{sec:sample_srccond},
we get 57,078 MRS spectra, which belong to 10,441 stellar sources.
This MRS time-domain sample of solar-like stars is employed for the following analysis. 
Table \ref{tab:counts_sample_selection} gives the counts of the selected MRS spectra and stellar sources after each of the screening conditions for sample selection.
The number in parentheses is the change in count relative to the previous step.
The bottom row of Table \ref{tab:counts_sample_selection} is also the counts of the final MRS time-domain sample of solar-like stars employed in this work after all the screening conditions.

\begin{table}[tbp]
  \centering
  \caption{Counts of MRS spectra and stellar sources after each of the screening conditions for sample selection.}
  \label{tab:counts_sample_selection}
  \begin{tabular*}{15.9cm}{@{\extracolsep\fill}rrcrrcl}
    \toprule
     \multicolumn{2}{r}{Count of MRS Spectra} && \multicolumn{2}{r}{Count of Stellar Sources} && Original MRS Catalog \\
    \hline
                     & 2,148,470 &&              & 1,019,861 && {\tt LAMOST MRS Parameter Catalog} (DR10 v1.0) \\
    \toprule
     \multicolumn{2}{r}{Count of MRS Spectra} && \multicolumn{2}{r}{Count of Stellar Sources} && Screening Condition for MRS Spectra \\
    \hline
      ($-$1,311,123) & 837,347   && ($-$608,417) & 411,444   &&  solar-like condition \\
      ($-$576,608)   & 260,739   && ($-$277,704) & 133,740   &&  signal-to-noise ratio condition \\
      ($-$69,496)    & 191,243   && ($-$29,936)  & 103,804   &&  removing low-quality spectra \\
      ($-$35,150)    & 156,093   && ($-$16,731)  &  87,073   &&  removing spectra with large projected rotational velocities \\
      ($-$5,964)     & 150,129   && ($-$2,902)   &  84,171   &&  removing spectra contaminated by the H\,II regions \\
    \toprule
     \multicolumn{2}{r}{Count of MRS Spectra} && \multicolumn{2}{r}{Count of Stellar Sources} && Screening Condition for Stellar Sources \\
    \hline
      ($-$81,645)    &  68,484   && ($-$72,030)  &  12,141   &&  time-domain condition \\
      ($-$8,218)     &  60,266   && ($-$1,247)   &  10,894   &&  condition for consistency of stellar atmospheric parameters \\
      ($-$3,145)     &  57,121   && ($-$447)     &  10,447   &&  condition for consistency of H$\alpha$ profiles \\
      ($-$43)        &  57,078   && ($-$6)       &  10,441   &&  removing sources in the MRS catalog of close binaries \\ 
    \botrule
  \end{tabular*}
  \footnotetext{
    Note:
    (1) The screening conditions are applied from top to bottom sequentially; 
    the bottom row is also the counts of the final MRS time-domain sample of solar-like stars employed in this work.
    (2) The number in parentheses is the change in count relative to the previous step.}
\end{table}

\begin{figure}
  \centering
  \includegraphics[width=0.53\textwidth]{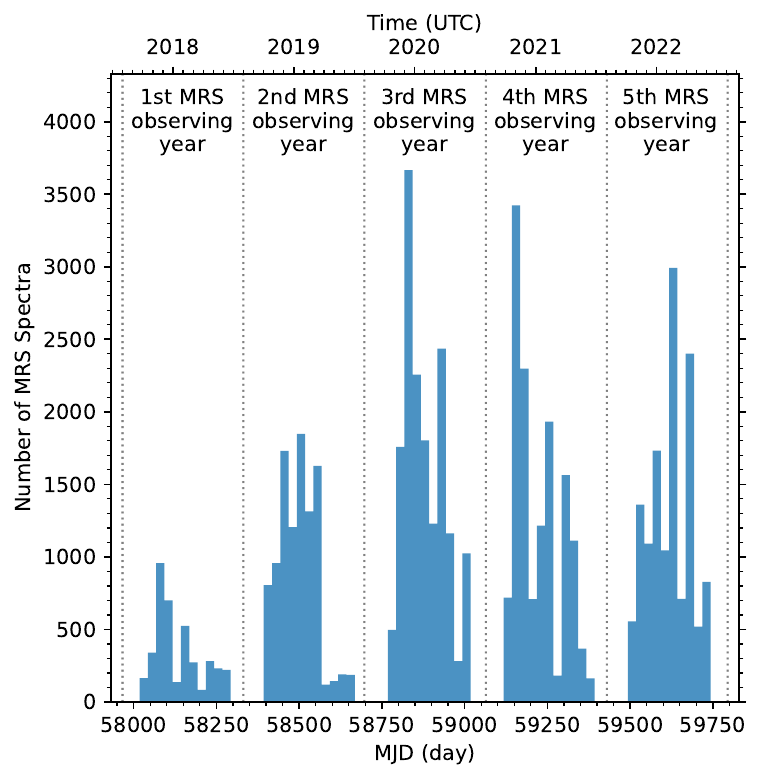}
  \caption{Observing timeline for all the spectra in the MRS time-domain sample of solar-like stars employed in this work.
    The bottom axis shows the observation time using the modified Julian day ($\mathrm{MJD} = \mathrm{JD} - 2400000.5$), and the top axis shows the UTC time of the observations.
  The vertical dotted lines (corresponding to August 1st 00:00 UTC of each calendar year) separate the five observing years of MRS.
  The width of each bin in the diagram is 25 days.}
  \label{fig:timeline_mrs_timedomain_spectra}
\end{figure}

Figure \ref{fig:timeline_mrs_timedomain_spectra} shows the observing timeline for all the spectra in the selected MRS time-domain sample of solar-like stars.
The dates of the spectra span five observing years of MRS (from September 2017 to June 2022).
The absence of observations in July and August of each calendar year (time slots indicated by the vertical dotted lines in Figure \ref{fig:timeline_mrs_timedomain_spectra}) is due to the weather condition and the maintenance of the telescope, 
which naturally separates the individual observing years. 

\begin{figure}
  \centering
  \includegraphics[width=0.455\textwidth]{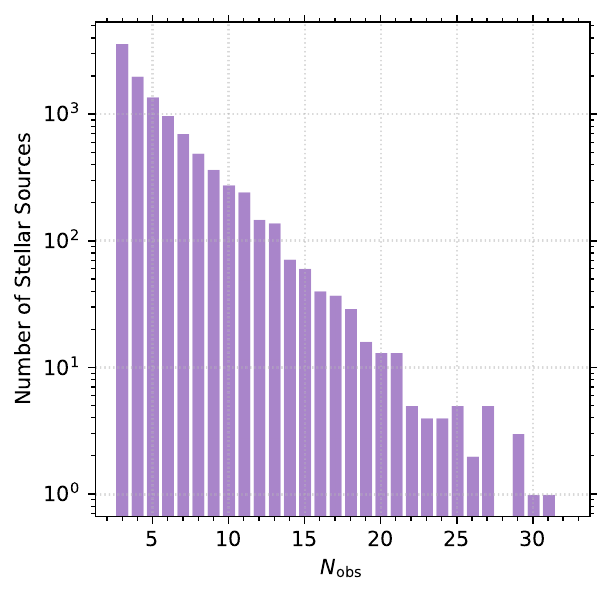}
  \caption{Distribution histogram of $N_\mathrm{obs}$ for all the stellar sources in the MRS time-domain sample of solar-like stars employed in this work.}
  \label{fig:numobs_mrs_timedomain_source}
\end{figure}

Figure \ref{fig:numobs_mrs_timedomain_source} shows the distribution histogram of $N_\mathrm{obs}$ (number of MRS observations of each stellar source; see definition in Section \ref{sec:sample_srccond_timedomain}) for all the stellar sources in the selected MRS time-domain sample of solar-like stars.
It can be seen from Figure \ref{fig:numobs_mrs_timedomain_source} that the value of $N_\mathrm{obs}$ in our sample varies in the range from $3$ to $31$,
and the number of stellar sources corresponding to each $N_\mathrm{obs}$ value has the trend of rapid decrease with the increase of $N_\mathrm{obs}$.

\begin{figure}
  \centering
  \includegraphics[width=0.60\textwidth]{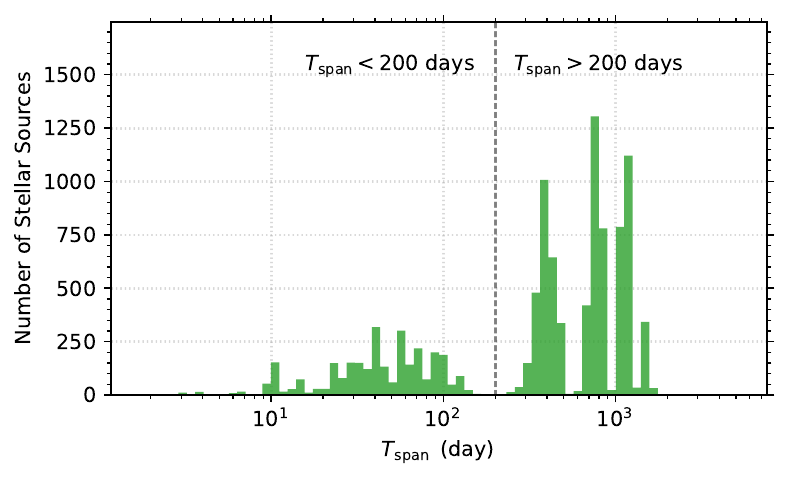}
  \caption{Distribution histogram of $T_\mathrm{span}$ for all the stellar sources in the MRS time-domain sample of solar-like stars employed in this work.
  The vertical dashed line corresponds to $T_\mathrm{span} = 200$~days, which divides the full sample of stellar sources into two groups ($T_\mathrm{span} < 200$~days and $T_\mathrm{span} > 200$~days). 
  }
  \label{fig:timespan_mrs_timedomain_source}
\end{figure}

Figure \ref{fig:timespan_mrs_timedomain_source} shows the distribution histogram of $T_\mathrm{span}$ (time span of multiple observations of each stellar source; see definition in Section \ref{sec:sample_srccond_timedomain}) for all the stellar sources in the selected MRS time-domain sample of solar-like stars.
It can be seen from Figure~\ref{fig:timespan_mrs_timedomain_source} that the full sample of stellar sources can be divided into two groups according to their $T_\mathrm{span}$ values.
One group has a time span within a few months ($T_\mathrm{span} < 200$\,days) and the other group has a time span 
of approximately one year or several years ($T_\mathrm{span} > 200$\,days), 
and the latter group is in the majority (the numbers of stellar sources of the two groups being 2,902 and 7,539, respectively).
The behaviors of the stellar sources with different $T_\mathrm{span}$ scales will be investigated in the following analysis.

The dataset of the selected MRS time-domain sample of solar-like stars employed in this work is available online.
Details of the dataset and the web link to the dataset can be found in Appendix~\ref{sec:appendix_dataset}.

In the following analysis, we mainly use the stellar atmospheric parameters of the source defined in Equation~(\ref{equ:def_source_paras}).
For concise expression, the subscript `source' is omitted.
If not specifically noted, the stellar atmospheric parameters and associated symbols refer to the values of the source.

\section{Data processing for H$\alpha$ variability} \label{sec:measure}

\subsection{Measure of H$\alpha$ activity intensity} \label{sec:measure_activityintensity}

We use the activity indices introduced for the MRS H$\alpha$ spectral data in \citet{2023Ap&SS.368...63H} to measure H$\alpha$ activity intensity.
The primary measure of H$\alpha$ activity intensity is the H$\alpha$ activity index (denoted by $I_\mathrm{H\alpha}$),
which is defined as the ratio of the mean flux in the central band of the H$\alpha$ line to the mean flux in the two continuum bands on the two sides of the line.
To evaluate $I_\mathrm{H\alpha}$ index, the width of the central band of the H$\alpha$ line in MRS data is set to 0.25\,{\AA},
and the two continuum bands are both 25\,{\AA} away from the H$\alpha$ line center and 5\,{\AA} in width \citep{2023Ap&SS.368...63H}.
The wavelength shift in the MRS spectral data caused by radial velocity is corrected by using the {\tt rv\_r0} parameter provided in the MRS catalog (see Section \ref{sec:sample_spcond_quality}).

Because the value of the $I_\mathrm{H\alpha}$ index depends on the continuum flux, 
which is a function of stellar atmospheric parameters,
the H$\alpha$ $R$-index (denoted by $R_\mathrm{H\alpha}$) is introduced as the secondary measure of H$\alpha$ activity intensity,
which is defined as the ratio of H$\alpha$ luminosity to bolometric luminosity.
The relationship between $R_\mathrm{H\alpha}$ index and $I_\mathrm{H\alpha}$ index can be described by Equation \citep{2023Ap&SS.368...63H}
\begin{equation} \label{equ:Rindex}
  R_\mathrm{H\alpha} = \frac{I_\mathrm{H\alpha} \cdot \overline{f}_\mathrm{cont} \cdot \Delta\lambda} 
  {\sigma T_\mathrm{eff}^4},
\end{equation}
where $\overline{f}_\mathrm{cont}$ is the mean of the stellar absolute spectral flux in the two continuum bands (derived from the PHOENIX stellar atmospheric model published in \citealt{2013A&A...553A...6H}),
and $\Delta\lambda$ is the width of the central band of H$\alpha$ line (0.25\,{\AA} as described above). 
Then the $I_\mathrm{H\alpha} \cdot \overline{f}_\mathrm{cont} \cdot \Delta\lambda$ term in Equation (\ref{equ:Rindex}) is the total absolute flux in the central band of the H$\alpha$ line,
which is divided by the bolometric flux $\sigma T_\mathrm{eff}^4$ (where $\sigma$ is the Stefan-Boltzmann constant) to yield the $R_\mathrm{H\alpha}$ index.

We evaluate the values of $I_\mathrm{H\alpha}$ and $R_\mathrm{H\alpha}$ indices for all the spectra in the MRS time-domain sample of solar-like stars employed in this work. 
The uncertainties of the $I_\mathrm{H\alpha}$ and $R_\mathrm{H\alpha}$ values (denoted by $\delta I_\mathrm{H\alpha}$ and $\delta R_\mathrm{H\alpha}$, respectively) are also estimated by using the approach described in \citet{2023Ap&SS.368...63H}.
In this work, when evaluating $R_\mathrm{H\alpha}$ index, 
the spectra of the same stellar source use the same stellar atmospheric parameters of the source defined in Equation~(\ref{equ:def_source_paras}) to maintain consistency between the $R_\mathrm{H\alpha}$ values of the co-source spectra.
In the work of \citet{2023Ap&SS.368...63H}, when calculating $R_\mathrm{H\alpha}$,
each spectrum uses its own stellar atmospheric parameter values provided in the MRS catalog,
which is different from the scheme adopted in this work.

Because the $R_\mathrm{H\alpha}$ index reflects the absolute flux in the central band of the H$\alpha$ line and hence is more physical than the $I_\mathrm{H\alpha}$ index,
the measure of H$\alpha$ variability described in the next subsection is based on the $R_\mathrm{H\alpha}$ index. 
More information and diagrams of the $I_\mathrm{H\alpha}$ index are provided in Appendix \ref{sec:appendix_iha}.
Both the $R_\mathrm{H\alpha}$ and $I_\mathrm{H\alpha}$ data of the MRS spectra employed in this work are available in the dataset of this work (see Appendix \ref{sec:appendix_dataset}).

For the convenience of following analysis, 
we assign one representative value of $R_\mathrm{H\alpha}$ to each stellar source in the MRS time-domain sample employed in this work by using the median of the multiple measurements of $R_\mathrm{H\alpha}$ (denoted by $R_\mathrm{H\alpha}^\mathrm{median}$). 
That is, for a stellar source, the representative value of $R_\mathrm{H\alpha}$ is
\begin{equation} \label{equ:source-Rhalpha}
  R_\mathrm{H\alpha}^\mathrm{median} = \mathrm{median}(\{R_\mathrm{H\alpha}^i\}), 
\end{equation}
where $\{R_\mathrm{H\alpha}^i, i=1, \cdots, N_\mathrm{obs}\}$ is the set of multiple $R_\mathrm{H\alpha}$ measurements of the source.
The data of $R_\mathrm{H\alpha}^\mathrm{median}$ are also available in the dataset of this work (see Appendix \ref{sec:appendix_dataset}).

Figure \ref{fig:rha_vs_teff_mrs_timedomain_spectra} shows the distributions of the derived $R_\mathrm{H\alpha}$ values with stellar effective temperature for the spectra (in Figure \ref{fig:rha_vs_teff_mrs_timedomain_spectra}a) and stellar sources (in Figure \ref{fig:rha_vs_teff_mrs_timedomain_spectra}b) in the MRS time-domain sample of solar-like stars employed in this work.
Each dot in Figure~\ref{fig:rha_vs_teff_mrs_timedomain_spectra}a corresponds to a MRS spectrum,
and the $R_\mathrm{H\alpha}$ values from the multiple observations of the same stellar source are connected by a vertical line segment;
as a reference, Figure \ref{fig:rha_vs_teff_mrs_timedomain_spectra}b displays the $R_\mathrm{H\alpha}$ distribution for the stellar sources by using the $R_\mathrm{H\alpha}^\mathrm{median}$ values (each dot corresponding to a stellar source).
As exhibited in Figure \ref{fig:rha_vs_teff_mrs_timedomain_spectra},
for the MRS time-domain sample employed in this work,
the $R_\mathrm{H\alpha}$ values is on the order of magnitude of $10^{-5}$.
The median of $\delta R_\mathrm{H\alpha}$ (uncertainties of the $R_\mathrm{H\alpha}$ values) is about $1.4 \times 10^{-7}$,
which is shown as an error bar ($\pm \delta R_\mathrm{H\alpha}$) in the bottom right corner of the diagrams in Figure \ref{fig:rha_vs_teff_mrs_timedomain_spectra}.

The $R_\mathrm{H\alpha}$ distributions with stellar atmospheric parameters (such as the $R_\mathrm{H\alpha}$ distribution with $T_\mathrm{eff}$ shown in Figure \ref{fig:rha_vs_teff_mrs_timedomain_spectra}) have been investigated in detail in \citet{2023Ap&SS.368...63H}.
In this work, we focus on the variability of the $R_\mathrm{H\alpha}$ values.
It can be seen from Figure \ref{fig:rha_vs_teff_mrs_timedomain_spectra}a that for each stellar source,
the values of the multiple $R_\mathrm{H\alpha}$ measurements are generally not a constant and fluctuate within a certain range.
This variability of the $R_\mathrm{H\alpha}$ values will be quantitatively evaluated in Section \ref{sec:measure_variability}.

\begin{figure}
  \centering
  \includegraphics[width=0.94\textwidth]{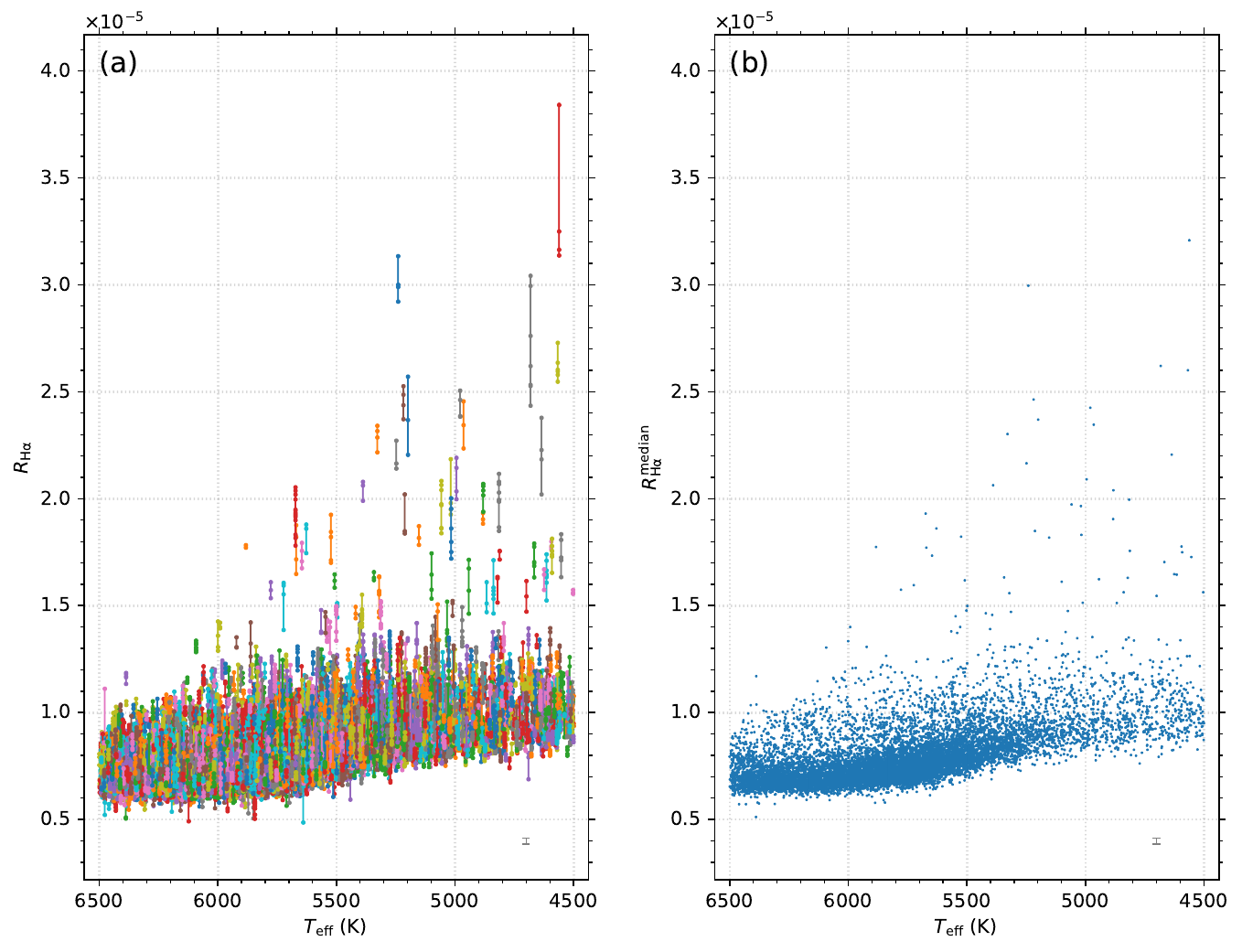}
  \caption{Distribution of $R_\mathrm{H\alpha}$ index with stellar effective temperature for (a) the spectra and (b) the stellar sources (using $R_\mathrm{H\alpha}^\mathrm{median}$) in the MRS time-domain sample of solar-like stars employed in this work. 
In panel (a), the $R_\mathrm{H\alpha}$ values from the multiple observations of the same stellar source are connected by a vertical line segment,
and the different line segments (stellar sources) are displayed in a variety of colors for ease of distinction.
The error bar ($\pm \delta R_\mathrm{H\alpha}$) in the bottom right corner of each panel shows the median of the uncertainties of the $R_\mathrm{H\alpha}$ values (about $1.4 \times 10^{-7}$).
   }
  \label{fig:rha_vs_teff_mrs_timedomain_spectra}
\end{figure}

\subsection{Measure of H$\alpha$ variability} \label{sec:measure_variability}

In the literature, the variability of stellar activity for a stellar source is often indicated by the STD (standard deviation) of the multiple measurements of an activity index (e.g., \citealt{2007ApJS..171..260L, 2009AJ....138..312H, 2021A&A...646A..77G}) as introduced in Section \ref{sec:intro}.
In this study, the $N_\mathrm{obs}$ of multiple measurements varies in the range from $3$ to $31$ (see Figure \ref{fig:numobs_mrs_timedomain_source}).
Because the definition of STD depends on $N_\mathrm{obs}$, the STD results corresponding to small and large $N_\mathrm{obs}$ values can not be compared directly.
For this reason, in this work we use extent of $R_\mathrm{H\alpha}$ fluctuation instead of STD to measure stellar H$\alpha$ variability. 

For the multiple $R_\mathrm{H\alpha}$ measurements of a stellar source, 
the extent of $R_\mathrm{H\alpha}$ fluctuation (denoted by $R_{\mathrm{H}{\alpha}}^\mathrm{EXT}$) is defined as the difference between the maximum and minimum $R_\mathrm{H\alpha}$ values of the source, i.e., 
\begin{equation} \label{equ:extrha}
  R_{\mathrm{H}{\alpha}}^\mathrm{EXT} = \max(\{R_\mathrm{H\alpha}^i\}) - \min(\{R_\mathrm{H\alpha}^i\}).
\end{equation}
The value of $R_{\mathrm{H}{\alpha}}^\mathrm{EXT}$ corresponds to the length of the line segments displayed in Figure~\ref{fig:rha_vs_teff_mrs_timedomain_spectra}a. 
The advantage of $R_{\mathrm{H}{\alpha}}^\mathrm{EXT}$ as a measure of H$\alpha$ variability is that its definition does not depend on $N_\mathrm{obs}$,
and hence the $R_{\mathrm{H}{\alpha}}^\mathrm{EXT}$ results associated with different $N_\mathrm{obs}$ values can be compared directly.
From a statistical point of view, large sample size of stellar sources can compensate for small $N_\mathrm{obs}$,
and this is the case in this study (see Figure \ref{fig:numobs_mrs_timedomain_source}).
We evaluate the $R_{\mathrm{H}{\alpha}}^\mathrm{EXT}$ values and their uncertainties (denoted by $\delta R_{\mathrm{H}{\alpha}}^\mathrm{EXT}$) based on Equation (\ref{equ:extrha}) for all the stellar sources in the MRS time-domain sample of solar-like stars employed in this work.
The STD of $R_\mathrm{H\alpha}$ fluctuation (denoted by $R_{\mathrm{H}{\alpha}}^\mathrm{STD}$) is also calculated for the sample,
and the relationship between the values of $R_{\mathrm{H}{\alpha}}^\mathrm{EXT}$ and $R_{\mathrm{H}{\alpha}}^\mathrm{STD}$ will be discussed in Section \ref{sec:discu}.
The data of the derived $R_{\mathrm{H}{\alpha}}^\mathrm{EXT}$ and $R_{\mathrm{H}{\alpha}}^\mathrm{STD}$ are available in the dataset of this work (see Appendix \ref{sec:appendix_dataset}).

\begin{figure}
  \centering
  \includegraphics[width=0.99\textwidth]{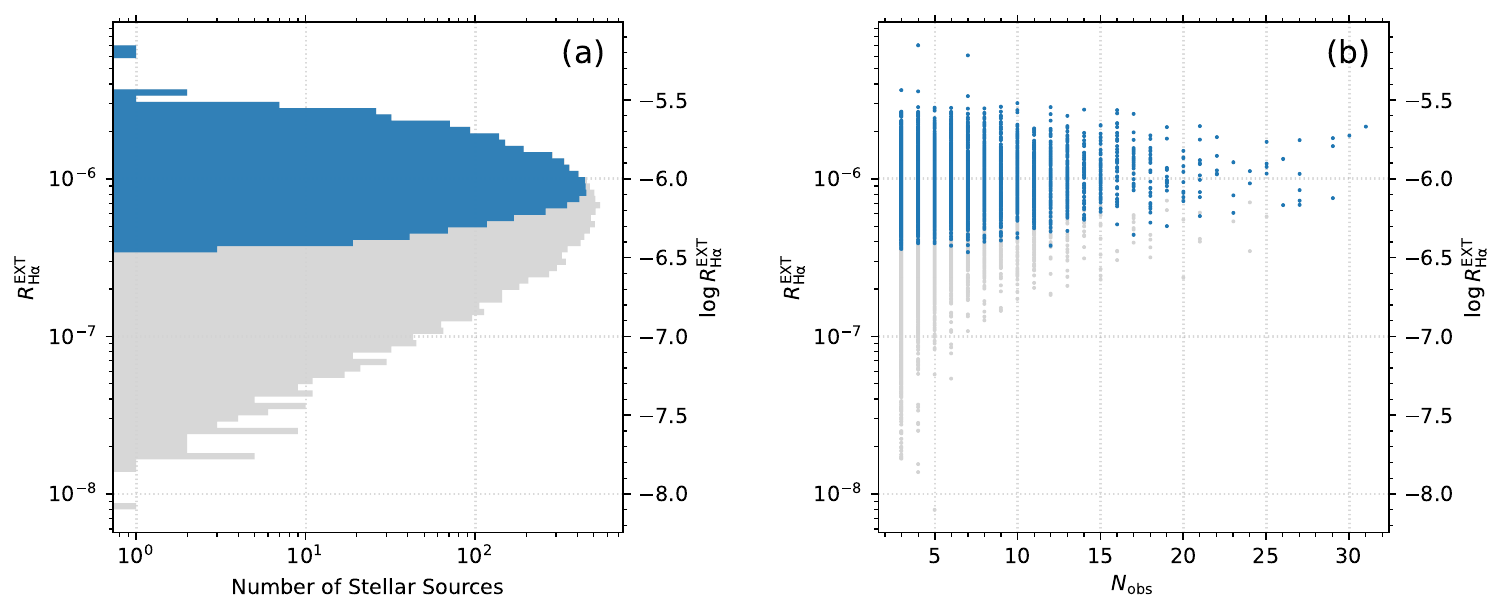}
  \caption{(a) Distribution histogram of $R_{\mathrm{H}{\alpha}}^\mathrm{EXT}$ for the stellar sources in the MRS time-domain sample of solar-like stars employed in this work.
  The histogram in light gray is for the full sample, 
  and the overlapping histogram in blue is for the sample of $R_{\mathrm{H}{\alpha}}^\mathrm{EXT} > 3 \times \delta R_{\mathrm{H}{\alpha}}^\mathrm{EXT}$. 
(b) Distribution of $R_{\mathrm{H}{\alpha}}^\mathrm{EXT}$ with $N_\mathrm{obs}$.
The data points in light gray is for the full sample,
and the overlapping data points in blue is for the sample of $R_{\mathrm{H}{\alpha}}^\mathrm{EXT} > 3 \times \delta R_{\mathrm{H}{\alpha}}^\mathrm{EXT}$. 
  }
  \label{fig:extrha_histogram_and_vs_numobs}
\end{figure}

Figure \ref{fig:extrha_histogram_and_vs_numobs}a shows the distribution histogram of the $R_{\mathrm{H}{\alpha}}^\mathrm{EXT}$ values for the stellar sources in the MRS time-domain sample of solar-like stars employed in this work, 
and Figure \ref{fig:extrha_histogram_and_vs_numobs}b shows the distribution of $R_{\mathrm{H}{\alpha}}^\mathrm{EXT}$ with $N_\mathrm{obs}$.
In Figure \ref{fig:extrha_histogram_and_vs_numobs}, the histogram and data points in light gray are for the full sample,
and the overlapping histogram and data points in blue are for the sample of $R_{\mathrm{H}{\alpha}}^\mathrm{EXT} > 3 \times \delta R_{\mathrm{H}{\alpha}}^\mathrm{EXT}$ (i.e., the sample with signals above the three-times-uncertainty threshold).
It can be seen from Figure \ref{fig:extrha_histogram_and_vs_numobs}a that the peak of the $R_{\mathrm{H}{\alpha}}^\mathrm{EXT}$ distribution is at about $10^{-6}$, 
which is one order of magnitude smaller than the values of $R_{\mathrm{H}{\alpha}}$ 
(about $10^{-5}$; see Section \ref{sec:measure_activityintensity}). 
It can also be seen from Figure~\ref{fig:extrha_histogram_and_vs_numobs}a that a considerable proportion of the stellar sources in our sample have the $R_{\mathrm{H}{\alpha}}^\mathrm{EXT}$ values below the $3 \times \delta R_{\mathrm{H}{\alpha}}^\mathrm{EXT}$ threshold (the median of $3 \times \delta R_{\mathrm{H}{\alpha}}^\mathrm{EXT}$ being about $6.2 \times 10^{-7}$),
and Figure \ref{fig:extrha_histogram_and_vs_numobs}b further indicates that this portion of the sample is generally associated with smaller $N_\mathrm{obs}$.
As $N_\mathrm{obs}$ increases, the random fluctuations between the multiple measurements bring the lower envelope of the $R_{\mathrm{H}{\alpha}}^\mathrm{EXT}$ values approach the threshold, as demonstrated in Figure \ref{fig:extrha_histogram_and_vs_numobs}b.
The stellar source samples with $R_{\mathrm{H}{\alpha}}^\mathrm{EXT} > 3 \times \delta R_{\mathrm{H}{\alpha}}^\mathrm{EXT}$ and $R_{\mathrm{H}{\alpha}}^\mathrm{EXT} \le 3 \times \delta R_{\mathrm{H}{\alpha}}^\mathrm{EXT}$ are distinguished in this work because the latter does not provide physical information on the distribution of $R_{\mathrm{H}{\alpha}}^\mathrm{EXT}$ 
(that is, the signals being eliminated by the noises),
yet it is still meaningful for the distribution of $R_{\mathrm{H}{\alpha}}$ as demonstrated in 
Figure \ref{fig:rha_vs_teff_mrs_timedomain_spectra}. 

Since the distribution range of the $R_{\mathrm{H}{\alpha}}^\mathrm{EXT}$ values is relatively large,
it is more convenient to use $\log R_{\mathrm{H}{\alpha}}^\mathrm{EXT}$ to express the distribution of $R_{\mathrm{H}{\alpha}}^\mathrm{EXT}$, 
as indicated by the right vertical axis of the diagrams in Figure \ref{fig:extrha_histogram_and_vs_numobs}.
Detailed analysis of the properties of the $\log R_{\mathrm{H}{\alpha}}^\mathrm{EXT}$ distribution for the MRS time-domain sample of solar-like stars employed in this work will be given in Section \ref{sec:result}.

\section{Variability of H$\alpha$ chromospheric activity of solar-like stars} \label{sec:result} 

As introduced in Section \ref{sec:intro}, one of the concerns about stellar activity variability is its relationship with stellar activity intensity,
so we first investigate the distribution of $\log R_{\mathrm{H}{\alpha}}^\mathrm{EXT}$ (representing H$\alpha$ variability) versus $\log R_{\mathrm{H}{\alpha}}^\mathrm{median}$ (representing H$\alpha$ activity intensity) in Section \ref{sec:result_logextrha_with_logmedianrha}.
Considering that the stellar sources in our sample have different values of $T_\mathrm{span}$ (time span of multiple observations) as illustrated in Figure \ref{fig:timespan_mrs_timedomain_source},
the distributions of $\log R_{\mathrm{H}{\alpha}}^\mathrm{EXT}$ versus $\log R_{\mathrm{H}{\alpha}}^\mathrm{median}$ for the stellar sources with different $T_\mathrm{span}$ scales are compared and investigated in Sections~\ref{sec:result_logextrha_with_tspan}.
We are also interested in the distribution of $\log R_{\mathrm{H}{\alpha}}^\mathrm{EXT}$ in the $T_\mathrm{eff}$ -- $\log R_{\mathrm{H}{\alpha}}^\mathrm{median}$ parameter space,
and this aspect is investigated in Section \ref{sec:result_logextrha_in_teff_logmedianrha_space}.
Because the stellar sources with $R_{\mathrm{H}{\alpha}}^\mathrm{EXT} \le 3 \times \delta R_{\mathrm{H}{\alpha}}^\mathrm{EXT}$ do not provide physical information on the distribution of $R_{\mathrm{H}{\alpha}}^\mathrm{EXT}$ (see Section \ref{sec:measure_variability}),
in the following analysis we only use the stellar sources with $R_{\mathrm{H}{\alpha}}^\mathrm{EXT} > 3 \times \delta R_{\mathrm{H}{\alpha}}^\mathrm{EXT}$ (4,877 in total).

\subsection{Distribution of $\log R_{\mathrm{H}{\alpha}}^\mathrm{EXT}$ versus $\log R_{\mathrm{H}{\alpha}}^\mathrm{median}$} \label{sec:result_logextrha_with_logmedianrha}

\begin{figure}
  \centering
  \includegraphics[width=0.67\textwidth]{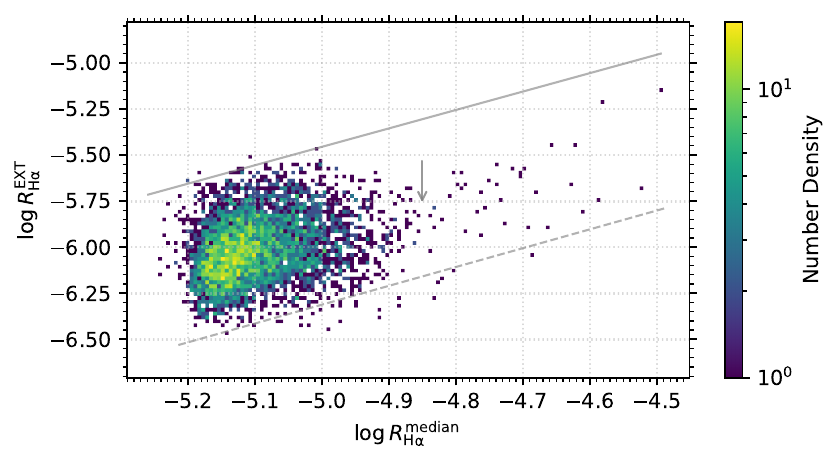}
  \caption{Distribution diagram of $\log R_{\mathrm{H}{\alpha}}^\mathrm{EXT}$ versus $\log R_{\mathrm{H}{\alpha}}^\mathrm{median}$ for the stellar sources in the MRS time-domain sample of solar-like stars employed in this work.
  Only the stellar sources meeting the $R_{\mathrm{H}{\alpha}}^\mathrm{EXT} > 3 \times \delta R_{\mathrm{H}{\alpha}}^\mathrm{EXT}$ condition are used in the diagram.
  Color scale indicates number density.
  The solid line and dashed line represent $R_{\mathrm{H}{\alpha}}^\mathrm{EXT} = 0.35 \times R_{\mathrm{H}{\alpha}}^\mathrm{median}$ and $R_{\mathrm{H}{\alpha}}^\mathrm{EXT} = 0.05 \times R_{\mathrm{H}{\alpha}}^\mathrm{median}$, respectively, 
  which enclose the area where the data points are mainly distributed.
  The short arrow indicates the turning position (corresponding to $\log R_{\mathrm{H}{\alpha}}^\mathrm{median} = -4.85$) of the top envelope of the data point distribution (see main text for details).
  }
  \label{fig:logextrha_vs_logmedianrha_density}
\end{figure}

In Figure \ref{fig:logextrha_vs_logmedianrha_density}, we show the distribution diagram of $\log R_{\mathrm{H}{\alpha}}^\mathrm{EXT}$ versus $\log R_{\mathrm{H}{\alpha}}^\mathrm{median}$ for the stellar sources in the MRS time-domain sample of solar-like stars employed in this work.
Only the stellar sources that meet the $R_{\mathrm{H}{\alpha}}^\mathrm{EXT} > 3 \times \delta R_{\mathrm{H}{\alpha}}^\mathrm{EXT}$ condition are used in Figure \ref{fig:logextrha_vs_logmedianrha_density}, as explained above. 
The number density of the data points is indicated by a color scale.
It can be seen from Figure \ref{fig:logextrha_vs_logmedianrha_density} that the data points are mainly distributed in the area between the lines of $R_{\mathrm{H}{\alpha}}^\mathrm{EXT} = 0.35 \times R_{\mathrm{H}{\alpha}}^\mathrm{median}$ and $R_{\mathrm{H}{\alpha}}^\mathrm{EXT} = 0.05 \times R_{\mathrm{H}{\alpha}}^\mathrm{median}$ (indicated by a solid line and a dashed line in Figure \ref{fig:logextrha_vs_logmedianrha_density}, respectively).
There is a positive correlation trend between $\log R_{\mathrm{H}{\alpha}}^\mathrm{EXT}$ and $\log R_{\mathrm{H}{\alpha}}^\mathrm{median}$ as exhibited by the data points in between the two lines, 
which is consistent with the results obtained by \citet{2007ApJS..171..260L} and \citet{2009AJ....138..312H} through the Ca II H and K lines (see introduction in Section \ref{sec:intro}).

The number density information in Figure \ref{fig:logextrha_vs_logmedianrha_density} shows that most of the data points are located in the lower left region of the diagram,
indicating that most stellar sources in our sample have low activity intensity and small H$\alpha$ variability. 
However, some stellar sources with lower activity intensity (in the left part of the diagram) do exhibit large H$\alpha$ variability (as large as reaching to the $R_{\mathrm{H}{\alpha}}^\mathrm{EXT} = 0.35 \times R_{\mathrm{H}{\alpha}}^\mathrm{median}$ line);
while for the stellar sources with higher activity intensity (in the right part of the diagram),
the data points are mainly distributed in the middle of the area enclosed by the $R_{\mathrm{H}{\alpha}}^\mathrm{EXT} = 0.35 \times R_{\mathrm{H}{\alpha}}^\mathrm{median}$ and $R_{\mathrm{H}{\alpha}}^\mathrm{EXT} = 0.05 \times R_{\mathrm{H}{\alpha}}^\mathrm{median}$ lines.
The result that solar-like stars with low activity intensity do not necessarily have small variability is also found by \citet{2021A&A...646A..77G} using the Ca II H and K lines, which is consistent with our result obtained from the H$\alpha$ line.

The lower envelope of the $\log R_{\mathrm{H}{\alpha}}^\mathrm{EXT}$ versus $\log R_{\mathrm{H}{\alpha}}^\mathrm{median}$ distribution (roughly coinciding with the $R_{\mathrm{H}{\alpha}}^\mathrm{EXT} = 0.05 \times R_{\mathrm{H}{\alpha}}^\mathrm{median}$ line in Figure \ref{fig:logextrha_vs_logmedianrha_density}) is influenced by the $R_{\mathrm{H}{\alpha}}^\mathrm{EXT} > 3 \times \delta R_{\mathrm{H}{\alpha}}^\mathrm{EXT}$ screening condition and mostly a truncation effect;
while the top envelope of the $\log R_{\mathrm{H}{\alpha}}^\mathrm{EXT}$ versus $\log R_{\mathrm{H}{\alpha}}^\mathrm{median}$ distribution exhibits fine structures,
which have physical implications.
As displayed in Figure \ref{fig:logextrha_vs_logmedianrha_density}, 
the top envelope of the $\log R_{\mathrm{H}{\alpha}}^\mathrm{EXT}$ versus $\log R_{\mathrm{H}{\alpha}}^\mathrm{median}$ distribution first increases with $\log R_{\mathrm{H}{\alpha}}^\mathrm{median}$,
reaching the $R_{\mathrm{H}{\alpha}}^\mathrm{EXT} = 0.35 \times R_{\mathrm{H}{\alpha}}^\mathrm{median}$ line around $\log R_{\mathrm{H}{\alpha}}^\mathrm{median} \sim -5.1$,
and then decreases with the increase of $\log R_{\mathrm{H}{\alpha}}^\mathrm{median}$ from about $\log R_{\mathrm{H}{\alpha}}^\mathrm{median} \sim -5.0$.
At about $\log R_{\mathrm{H}{\alpha}}^\mathrm{median} = -4.85$ (indicated by a short arrow in Figure \ref{fig:logextrha_vs_logmedianrha_density}), 
the top envelope turns to gradually increase with $\log R_{\mathrm{H}{\alpha}}^\mathrm{median}$,
and it is largely along a positive correlation line for $\log R_{\mathrm{H}{\alpha}}^\mathrm{median} > -4.85$. 
The different behaviors of the top envelope on the left side and right side of the turning position (short arrow) might be related to the stratification of activity levels of solar-like stars,
which will be further discussed in Section~\ref{sec:discu}.  

In the following analysis, the stellar sources on the left side ($\log R_{\mathrm{H}{\alpha}}^\mathrm{median} < -4.85$; with lower activity intensity) and right side ($\log R_{\mathrm{H}{\alpha}}^\mathrm{median} > -4.85$; with higher activity intensity) of the top envelope turning position will be distinguished where appropriate, owing to their different behaviors of H$\alpha$ variability.

\subsection{Distribution of $\log R_{\mathrm{H}{\alpha}}^\mathrm{EXT}$ versus $\log R_{\mathrm{H}{\alpha}}^\mathrm{median}$ for different $T_\mathrm{span}$ scales} \label{sec:result_logextrha_with_tspan}

\begin{figure}
  \centering
  \includegraphics[width=0.67\textwidth]{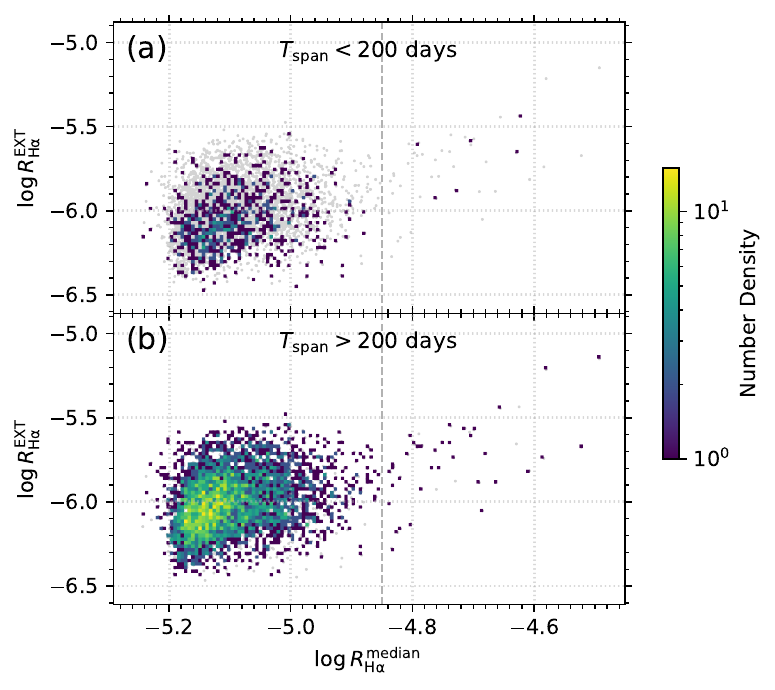}
  \caption{Distribution diagrams of $\log R_{\mathrm{H}{\alpha}}^\mathrm{EXT}$ versus $\log R_{\mathrm{H}{\alpha}}^\mathrm{median}$ for the stellar sources with (a) $T_\mathrm{span}<200$~days and (b) $T_\mathrm{span}>200$~days.
  Color scale indicates number density.
  The data points of the whole sample (as employed in Figure \ref{fig:logextrha_vs_logmedianrha_density}) are displayed in light gray in the background for reference.
  The vertical dashed line in each panel corresponds to $\log R_{\mathrm{H}{\alpha}}^\mathrm{median} = -4.85$, 
  which separates the stellar sources with lower ($\log R_{\mathrm{H}{\alpha}}^\mathrm{median} < -4.85$) and higher ($\log R_{\mathrm{H}{\alpha}}^\mathrm{median} > -4.85$) activity intensity. 
  }
  \label{fig:logextrha_vs_logmedianrha_tspan}
\end{figure}

In order to investigate how the time span of multiple observations ($T_\mathrm{span}$) affects the distribution of $\log R_{\mathrm{H}{\alpha}}^\mathrm{EXT}$ versus $\log R_{\mathrm{H}{\alpha}}^\mathrm{median}$,
in Figure \ref{fig:logextrha_vs_logmedianrha_tspan} we show the distribution diagrams of $\log R_{\mathrm{H}{\alpha}}^\mathrm{EXT}$ versus $\log R_{\mathrm{H}{\alpha}}^\mathrm{median}$ for the stellar sources with different $T_\mathrm{span}$ scales.
Figure \ref{fig:logextrha_vs_logmedianrha_tspan}a is for the stellar sources with $T_\mathrm{span} < 200$ days (within a few months),
and Figure \ref{fig:logextrha_vs_logmedianrha_tspan}b is for the stellar sources with $T_\mathrm{span} > 200$~days (approximately one year or several years).
The two groups of stellar sources with shorter and longer $T_\mathrm{span}$ scales have been described in Section~\ref{sec:sample_statistics}.
After the $R_{\mathrm{H}{\alpha}}^\mathrm{EXT} > 3 \times \delta R_{\mathrm{H}{\alpha}}^\mathrm{EXT}$ screening condition,
the numbers of stellar sources in the two groups are 757 and 4,120, respectively, as shown in Figures \ref{fig:logextrha_vs_logmedianrha_tspan}a and \ref{fig:logextrha_vs_logmedianrha_tspan}b.

Because the stellar sources with lower ($\log R_{\mathrm{H}{\alpha}}^\mathrm{median} < -4.85$) and higher ($\log R_{\mathrm{H}{\alpha}}^\mathrm{median} > -4.85$) activity intensity exhibit different behaviors of H$\alpha$ variability as described in Section \ref{sec:result_logextrha_with_logmedianrha},
we distinguish the two categories of stellar sources in the analysis, 
which are separated with a vertical dashed line (corresponds to $\log R_{\mathrm{H}{\alpha}}^\mathrm{median} = -4.85$) in each panel of Figure \ref{fig:logextrha_vs_logmedianrha_tspan}.

\begin{figure*}
  \centering
  \includegraphics[width=0.96\textwidth]{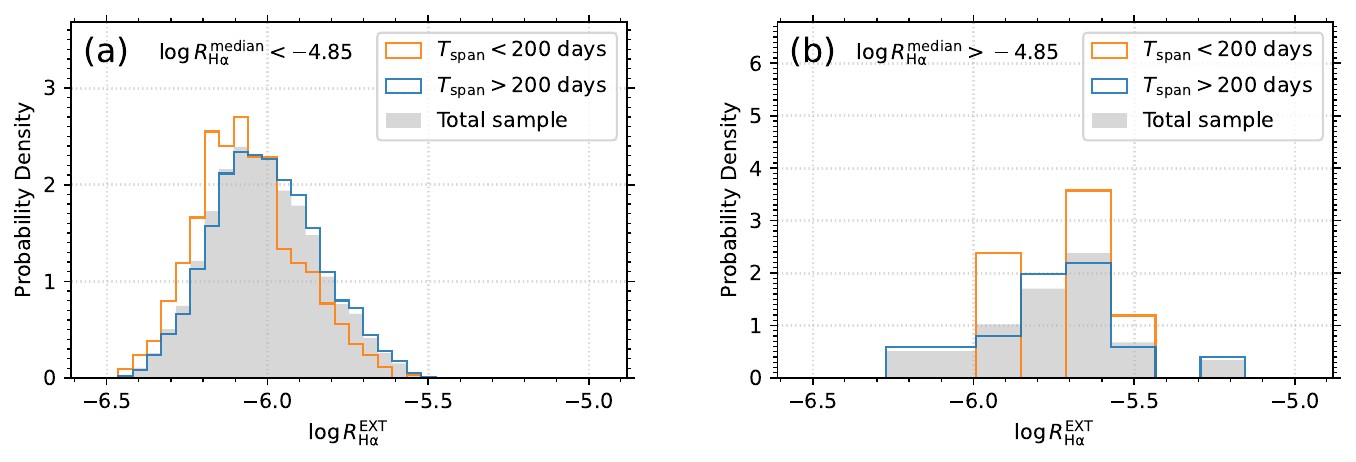}
  \caption{Distribution histograms of $\log R_{\mathrm{H}{\alpha}}^\mathrm{EXT}$ for the stellar source categories with (a) $\log R_{\mathrm{H}{\alpha}}^\mathrm{median} < -4.85$ (lower activity intensity) and (b) $\log R_{\mathrm{H}{\alpha}}^\mathrm{median} > -4.85$ (higher activity intensity).   
  In each panel, the histograms of the subsample with $T_\mathrm{span} < 200$~days, the subsample with $T_\mathrm{span} > 200$ days, and the total sample ($T_\mathrm{span} < 200$ days subsample plus $T_\mathrm{span} > 200$ days subsample) are displayed in orange line, blue line, and fill gray, respectively.
}
  \label{fig:logextrha_histogram_tspan}
\end{figure*}

It can be seen from Figure \ref{fig:logextrha_vs_logmedianrha_tspan}a that,
for the stellar sources with $T_\mathrm{span} < 200$ days,
the distribution of $\log R_{\mathrm{H}{\alpha}}^\mathrm{EXT}$ versus $\log R_{\mathrm{H}{\alpha}}^\mathrm{median}$ cannot reproduce the distribution configuration of the whole sample as shown in Figure~\ref{fig:logextrha_vs_logmedianrha_density},
especially for the data points with lower activity intensity (in the left part of the diagram) and in the zone near the top envelope;
while for the stellar sources with $T_\mathrm{span} > 200$~days (Figure \ref{fig:logextrha_vs_logmedianrha_tspan}b),
the distribution of $\log R_{\mathrm{H}{\alpha}}^\mathrm{EXT}$ versus $\log R_{\mathrm{H}{\alpha}}^\mathrm{median}$ is almost the same as the distribution configuration of the whole sample.
To confirm this impression, in Figure \ref{fig:logextrha_histogram_tspan} we show the distribution histograms of $\log R_{\mathrm{H}{\alpha}}^\mathrm{EXT}$,
and the $\log R_{\mathrm{H}{\alpha}}^\mathrm{EXT}$ histograms for the stellar source categories with $\log R_{\mathrm{H}{\alpha}}^\mathrm{median} < -4.85$ and $\log R_{\mathrm{H}{\alpha}}^\mathrm{median} > -4.85$ are given in panel (a) and panel (b) of Figure \ref{fig:logextrha_histogram_tspan}, respectively.
In each panel of Figure~\ref{fig:logextrha_histogram_tspan}, 
the $\log R_{\mathrm{H}{\alpha}}^\mathrm{EXT}$ histograms of the subsamples with $T_\mathrm{span} < 200$~days and $T_\mathrm{span} > 200$ days are displayed in orange line and blue line, respectively, 
and the histogram of the total sample ($T_\mathrm{span} < 200$ days subsample plus $T_\mathrm{span} > 200$ days subsample) is displayed in fill gray.
It can be seen from Figure \ref{fig:logextrha_histogram_tspan}a that for the stellar source category with $\log R_{\mathrm{H}{\alpha}}^\mathrm{median} < -4.85$,
the $\log R_{\mathrm{H}{\alpha}}^\mathrm{EXT}$ histogram of the subsample with $T_\mathrm{span} > 200$ days is almost the same as the histogram of the total sample,
while the histogram of the subsample with $T_\mathrm{span} < 200$ days shows leftward shift (direction of smaller $\log R_{\mathrm{H}{\alpha}}^\mathrm{EXT}$) relative to the histogram of the total sample,
which confirms the impression obtained from Figure \ref{fig:logextrha_vs_logmedianrha_tspan}.
This distinction of $\log R_{\mathrm{H}{\alpha}}^\mathrm{EXT}$ distribution for different $T_\mathrm{span}$ scales is not exhibited by the stellar source category with $\log R_{\mathrm{H}{\alpha}}^\mathrm{median} > -4.85$, as shown in Figure \ref{fig:logextrha_histogram_tspan}b.

The above result indicates that for the stellar sources with lower activity intensity ($\log R_{\mathrm{H}{\alpha}}^\mathrm{median} < -4.85$), the large $\log R_{\mathrm{H}{\alpha}}^\mathrm{EXT}$ values near the top envelope of the $\log R_{\mathrm{H}{\alpha}}^\mathrm{EXT}$ versus $\log R_{\mathrm{H}{\alpha}}^\mathrm{median}$ distribution tend to be associated with the long-term ($T_\mathrm{span} > 200$~days) variations of H$\alpha$ activity, 
which might be related to the stellar activity cycles introduced in Section \ref{sec:intro};
and the smaller $\log R_{\mathrm{H}{\alpha}}^\mathrm{EXT}$ values tend to be associated with the short-term ($T_\mathrm{span} < 200$~days) variations of H$\alpha$ activity, 
which might be related to the stellar rotational modulation introduced in Section \ref{sec:intro}. 
On the other hand, Figure~\ref{fig:logextrha_vs_logmedianrha_tspan}b shows that even for the sample with $T_\mathrm{span} > 200$~days,
a large part of the data points is still located in the lower left region of the diagram. 
This fact suggests that for this portion of stellar sources, 
if there are long-term trends of H$\alpha$ activity,
the amplitude of the long-term variations must be lower.
As for the stellar sources with higher activity intensity ($\log R_{\mathrm{H}{\alpha}}^\mathrm{median} > -4.85$),
there is no evidence of distinct distribution configurations between the long-term and short-term variations of H$\alpha$ activity, as shown in Figures \ref{fig:logextrha_vs_logmedianrha_tspan} and \ref{fig:logextrha_histogram_tspan}b.

\citet{2022A&A...658A..57M} analyzed the long-term and short-term variability of stellar H$\alpha$ and Ca II H and K activity by using the HARPS spectral data.
Most of the stellar objects in their sample are old solar-type main-sequence stars.
Their result shows that the distribution range of the short-term variability is smaller than that of the long-term variability, which is consistent with our result displayed in Figure \ref{fig:logextrha_histogram_tspan}a.

\subsection{Distribution of $\log R_{\mathrm{H}{\alpha}}^\mathrm{EXT}$ in $T_\mathrm{eff}$ -- $\log R_{\mathrm{H}{\alpha}}^\mathrm{median}$ parameter space} \label{sec:result_logextrha_in_teff_logmedianrha_space}

\begin{figure}
  \centering
  \includegraphics[width=0.60\textwidth]{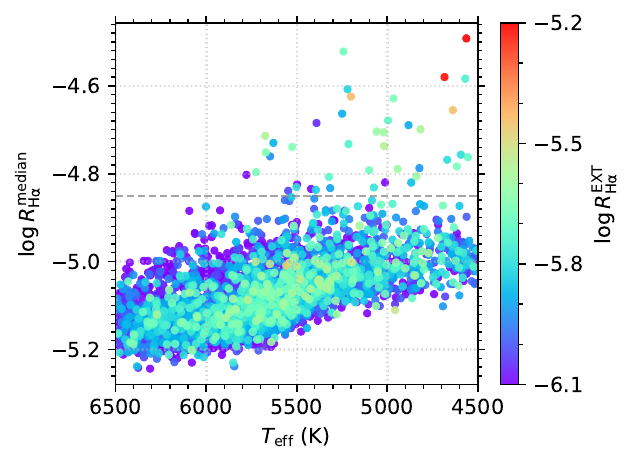}
  \caption{Distribution diagram of $\log R_{\mathrm{H}{\alpha}}^\mathrm{EXT}$ in the $T_\mathrm{eff}$ -- $\log R_{\mathrm{H}{\alpha}}^\mathrm{median}$ parameter space for the stellar sources in the MRS time-domain sample of solar-like stars employed in this work (same stellar source sample as employed in Figure \ref{fig:logextrha_vs_logmedianrha_density}). 
  The data points are sorted by their $\log R_{\mathrm{H}{\alpha}}^\mathrm{EXT}$ values, with the larger $\log R_{\mathrm{H}{\alpha}}^\mathrm{EXT}$ data points being displayed on top of the smaller $\log R_{\mathrm{H}{\alpha}}^\mathrm{EXT}$ data points.
  The value of $\log R_{\mathrm{H}{\alpha}}^\mathrm{EXT}$ of the data points is indicated by a color scale.
  The horizontal dashed line corresponds to $\log R_{\mathrm{H}{\alpha}}^\mathrm{median} = -4.85$,
  which separates the lower part ($\log R_{\mathrm{H}{\alpha}}^\mathrm{median} < -4.85$; with lower activity intensity) and upper part ($\log R_{\mathrm{H}{\alpha}}^\mathrm{median} > -4.85$; with higher activity intensity) of the diagram (see main text for details).
   }
  \label{fig:logextrha_in_teff_logmedianrha_space}
\end{figure}

In Figure \ref{fig:logextrha_in_teff_logmedianrha_space}, we show the distribution diagram of $\log R_{\mathrm{H}{\alpha}}^\mathrm{EXT}$ in the $T_\mathrm{eff}$ -- $\log R_{\mathrm{H}{\alpha}}^\mathrm{median}$ parameter space for the stellar sources in the MRS time-domain sample of solar-like stars employed in this work (same stellar source sample as employed in Figure \ref{fig:logextrha_vs_logmedianrha_density}).
The data points in Figure~\ref{fig:logextrha_in_teff_logmedianrha_space} are sorted by their $\log R_{\mathrm{H}{\alpha}}^\mathrm{EXT}$ values, with the larger $\log R_{\mathrm{H}{\alpha}}^\mathrm{EXT}$ data points being displayed on top of the smaller $\log R_{\mathrm{H}{\alpha}}^\mathrm{EXT}$ data points.
The value of $\log R_{\mathrm{H}{\alpha}}^\mathrm{EXT}$ of the data points is indicated by a color scale,
with $\log R_{\mathrm{H}{\alpha}}^\mathrm{EXT}$ less than $-5.8$ in blue, from $-5.8$ to $-5.5$ in green, and greater than $-5.5$ in red.
Note that the range of the $\log R_{\mathrm{H}{\alpha}}^\mathrm{EXT}$ values represented by green color ($\log R_{\mathrm{H}{\alpha}}^\mathrm{EXT} \sim -5.8$ to $-5.5$) corresponds to the top envelope of the $\log R_{\mathrm{H}{\alpha}}^\mathrm{EXT}$ versus $\log R_{\mathrm{H}{\alpha}}^\mathrm{median}$ distribution of the stellar source category with lower activity intensity ($\log R_{\mathrm{H}{\alpha}}^\mathrm{median} < -4.85$) described in Section \ref{sec:result_logextrha_with_logmedianrha}.

It can be seen from Figure \ref{fig:logextrha_in_teff_logmedianrha_space} that the distribution of $\log R_{\mathrm{H}{\alpha}}^\mathrm{EXT}$ in the $T_\mathrm{eff}$ -- $\log R_{\mathrm{H}{\alpha}}^\mathrm{median}$ parameter space has different properties in the lower and upper parts of the diagram,
which are separated by the $\log R_{\mathrm{H}{\alpha}}^\mathrm{median} = -4.85$ line (horizontal dashed line in Figure \ref{fig:logextrha_in_teff_logmedianrha_space}) and correspond to the stellar source categories with lower and higher activity intensity described in Section \ref{sec:result_logextrha_with_logmedianrha}.
In the lower part of the diagram ($\log R_{\mathrm{H}{\alpha}}^\mathrm{median} < -4.85$; with lower activity intensity),
the large $\log R_{\mathrm{H}{\alpha}}^\mathrm{EXT}$ values (in green) associated with the top envelope of the $\log R_{\mathrm{H}{\alpha}}^\mathrm{EXT}$ versus $\log R_{\mathrm{H}{\alpha}}^\mathrm{median}$ distribution are along a strip-shaped area near the bottom edge of the data point distribution, with the peak at about $T_\mathrm{eff} \sim 5600$\,K.
In the upper part of the diagram ($\log R_{\mathrm{H}{\alpha}}^\mathrm{median} > -4.85$; with higher activity intensity),
the largest $\log R_{\mathrm{H}{\alpha}}^\mathrm{EXT}$ values (in red) are located in the top right corner of the diagram and correspond to the smallest $T_\mathrm{eff}$ in our sample. 
Between the lower and upper parts of the diagram (around $\log R_{\mathrm{H}{\alpha}}^\mathrm{median} = -4.85$) is a valley of relatively small $\log R_{\mathrm{H}{\alpha}}^\mathrm{EXT}$ values,
which is also exhibited in Figure \ref{fig:logextrha_vs_logmedianrha_density}.
The physical implications of the different properties of the $\log R_{\mathrm{H}{\alpha}}^\mathrm{EXT}$ distribution for the stellar sources with lower and higher activity intensity will be discussed in Section \ref{sec:discu}.

\section{Discussion} \label{sec:discu}

\subsection{Relationship between $R_{\mathrm{H}{\alpha}}^\mathrm{EXT}$ and $R_{\mathrm{H}{\alpha}}^\mathrm{STD}$} \label{sec:discu_extrha_stdrha}

\begin{figure}
  \centering
  \includegraphics[width=0.48\textwidth]{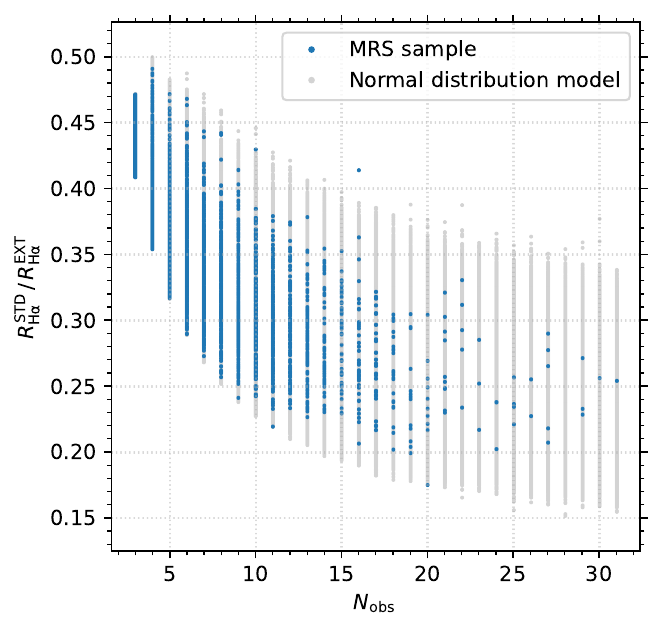}
  \caption{Distribution of $R_{\mathrm{H}{\alpha}}^\mathrm{STD} / R_{\mathrm{H}{\alpha}}^\mathrm{EXT}$ with $N_\mathrm{obs}$ for the stellar sources in the MRS time-domain sample of solar-like stars employed in this work (data points in blue; same stellar source sample as employed in Figure \ref{fig:logextrha_vs_logmedianrha_density}).
  The simulated $R_{\mathrm{H}{\alpha}}^\mathrm{STD} / R_{\mathrm{H}{\alpha}}^\mathrm{EXT}$ distribution based on a normal distribution model is displayed in light gray in the background for reference.
  }
  \label{fig:ratiostdextrha_vs_numobs}
\end{figure}

In the previous analysis, we use extent of $R_\mathrm{H\alpha}$ fluctuation ($R_{\mathrm{H}{\alpha}}^\mathrm{EXT}$) to measure H$\alpha$ variability for the MRS time-domain sample of solar-like stars employed in this work.
The STD of $R_\mathrm{H\alpha}$ fluctuation ($R_{\mathrm{H}{\alpha}}^\mathrm{STD}$) is also calculated for the sample since it is commonly adopted in the literature. 
In order to examine the relationship between the values of $R_{\mathrm{H}{\alpha}}^\mathrm{EXT}$ and $R_{\mathrm{H}{\alpha}}^\mathrm{STD}$,
in Figure \ref{fig:ratiostdextrha_vs_numobs} we show the distribution of the ratio value of $R_{\mathrm{H}{\alpha}}^\mathrm{STD} / R_{\mathrm{H}{\alpha}}^\mathrm{EXT}$ with $N_\mathrm{obs}$ for the stellar sources in the MRS time-domain sample of solar-like stars employed in this work (data points in blue; same stellar source sample as employed in Figure \ref{fig:logextrha_vs_logmedianrha_density}).
As a reference, we also simulate the $R_{\mathrm{H}{\alpha}}^\mathrm{STD} / R_{\mathrm{H}{\alpha}}^\mathrm{EXT}$ distribution by using a normal distribution model and through the Monte Carlo approach
(the sample size of the simulated distribution being $10^5$ for each of the $N_\mathrm{obs}$ values).
The simulated model distribution of $R_{\mathrm{H}{\alpha}}^\mathrm{STD} / R_{\mathrm{H}{\alpha}}^\mathrm{EXT}$ is displayed in light gray in the background in Figure \ref{fig:ratiostdextrha_vs_numobs}.

It can be seen from Figure \ref{fig:ratiostdextrha_vs_numobs} that the values of $R_{\mathrm{H}{\alpha}}^\mathrm{STD} / R_{\mathrm{H}{\alpha}}^\mathrm{EXT}$ of the normal distribution model (data points in light gray) have a decreasing trend with the increase of $N_\mathrm{obs}$,
and vary from about $0.50$ (corresponding to smaller $N_\mathrm{obs}$) to about $0.15$ (corresponding to larger $N_\mathrm{obs}$).
The distribution of the $R_{\mathrm{H}{\alpha}}^\mathrm{STD} / R_{\mathrm{H}{\alpha}}^\mathrm{EXT}$ values of the MRS sample (data points in blue) is generally consistent with the simulated distribution of the normal distribution model.

\begin{figure*}
  \centering
  \includegraphics[width=1.25\textwidth]{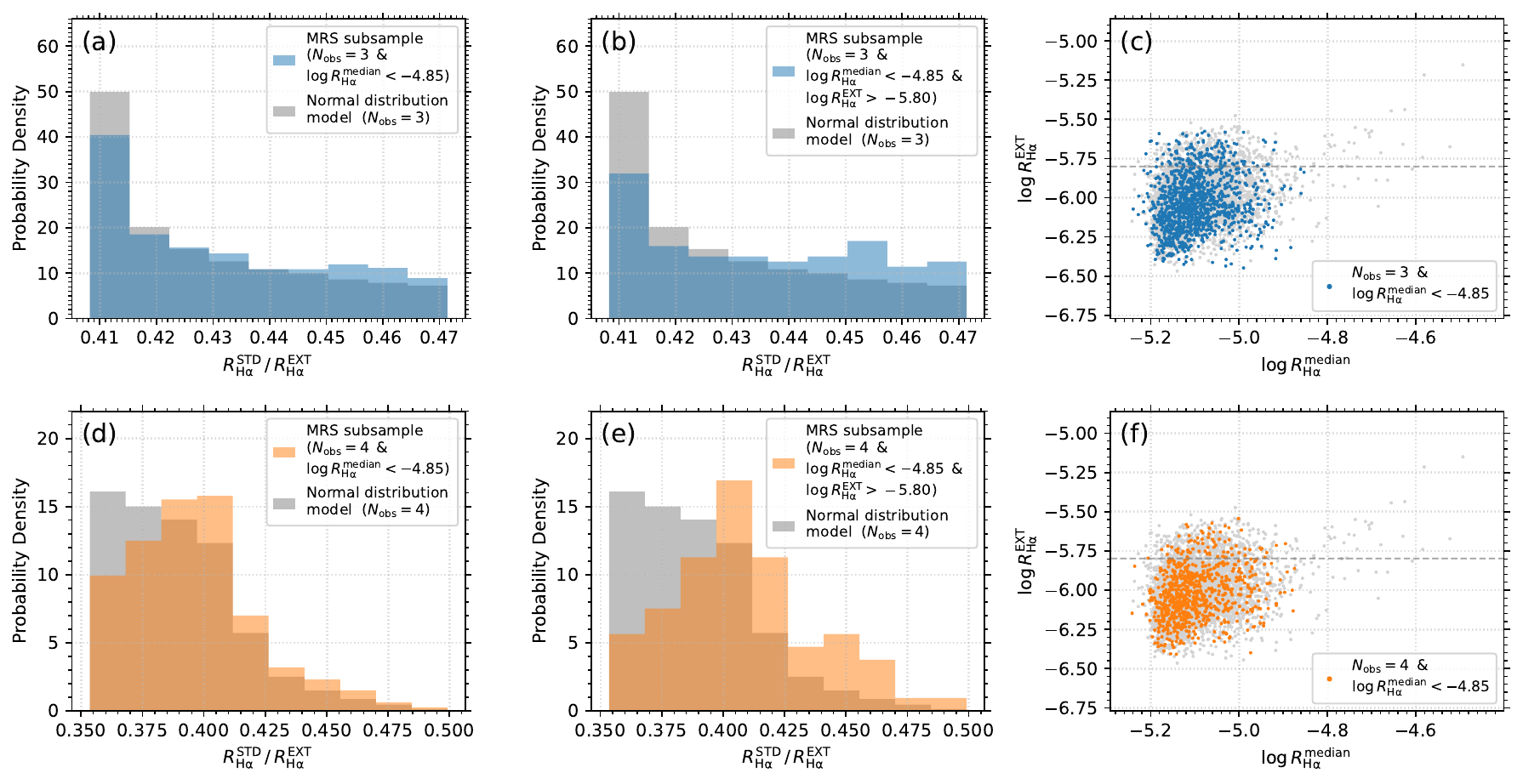}
  \caption{(leftmost column) 
  Histograms of $R_{\mathrm{H}{\alpha}}^\mathrm{STD} / R_{\mathrm{H}{\alpha}}^\mathrm{EXT}$ for the MRS subsample of stellar sources with $N_\mathrm{obs} = 3$ \& $\log R_\mathrm{H\alpha}^\mathrm{median} < -4.85$ (top panel; in blue) and the MRS subsample with $N_\mathrm{obs} = 4$ \& $\log R_\mathrm{H\alpha}^\mathrm{median} < -4.85$ (bottom panel; in orange), 
  compared with the simulated histograms of the normal distribution model (in gray). 
  (middle column) Same as in the leftmost column, but for the MRS subsamples after adding screening condition of $\log R_\mathrm{H\alpha}^\mathrm{EXT} > -5.80$.
  (rightmost column) 
  Scatter plots of $\log R_\mathrm{H\alpha}^\mathrm{EXT}$ versus $\log R_\mathrm{H\alpha}^\mathrm{median}$ for the two MRS subsamples employed in the leftmost column.
  The whole MRS stellar source sample (as employed in Figure \ref{fig:logextrha_vs_logmedianrha_density}) is displayed in light gray in the background for reference.
  The horizontal dashed lines correspond to $\log R_\mathrm{H\alpha}^\mathrm{EXT} = -5.80$; 
  the data points above the lines are used in the middle column.
  }
  \label{fig:ratiostdextrha_histogram}
\end{figure*}

In order to examine the degree of consistency and possible discrepancies between the $R_{\mathrm{H}{\alpha}}^\mathrm{STD} / R_{\mathrm{H}{\alpha}}^\mathrm{EXT}$ distributions of the MRS sample and the normal distribution model,
in Figures \ref{fig:ratiostdextrha_histogram}a and \ref{fig:ratiostdextrha_histogram}d (leftmost column of Figure \ref{fig:ratiostdextrha_histogram}), 
we show the histograms of $R_{\mathrm{H}{\alpha}}^\mathrm{STD} / R_{\mathrm{H}{\alpha}}^\mathrm{EXT}$ for the MRS subsample of stellar sources with $N_\mathrm{obs} = 3$ \& $\log R_\mathrm{H\alpha}^\mathrm{median} < -4.85$ (in blue) and the MRS subsample with $N_\mathrm{obs} = 4$ \& $\log R_\mathrm{H\alpha}^\mathrm{median} < -4.85$ (in orange), respectively, 
compared with the simulated histograms of the normal distribution model (in gray). 
The two $N_\mathrm{obs}$ values are employed because compared with the MRS subsamples with other $N_\mathrm{obs}$ values, 
the MRS subsamples with $N_\mathrm{obs} = 3$ and $4$ have a larger sample size (see Figure \ref{fig:numobs_mrs_timedomain_source}) and hence a higher level of statistical significance. 
The scatter plots of $\log R_\mathrm{H\alpha}^\mathrm{EXT}$ versus $\log R_\mathrm{H\alpha}^\mathrm{median}$ of the two MRS subsamples are given in Figures~\ref{fig:ratiostdextrha_histogram}c and \ref{fig:ratiostdextrha_histogram}f (rightmost column of Figure \ref{fig:ratiostdextrha_histogram}), respectively, for reference.
In Figure \ref{fig:ratiostdextrha_histogram}, we only use the MRS stellar sources with $\log R_\mathrm{H\alpha}^\mathrm{median} < -4.85$ (data points in the left part of the $\log R_\mathrm{H\alpha}^\mathrm{EXT}$ versus $\log R_\mathrm{H\alpha}^\mathrm{median}$ diagram; see Figures \ref{fig:ratiostdextrha_histogram}c and~\ref{fig:ratiostdextrha_histogram}f) because the stellar sources with $\log R_\mathrm{H\alpha}^\mathrm{median} < -4.85$ and $\log R_\mathrm{H\alpha}^\mathrm{median} > -4.85$ present different behaviors of H$\alpha$ variability as described in Section \ref{sec:result_logextrha_with_logmedianrha} and hence are distinguished in the analysis.
Since the method used in Figure \ref{fig:ratiostdextrha_histogram} cannot be applied only to the $\log R_\mathrm{H\alpha}^\mathrm{median} > -4.85$ stellar sources owing to their small sample size,
in Appendix \ref{sec:appendix_histratiostdextrha_wholesample}, we give the $R_{\mathrm{H}{\alpha}}^\mathrm{STD} / R_{\mathrm{H}{\alpha}}^\mathrm{EXT}$ histograms of $N_\mathrm{obs} = 3$ and $N_\mathrm{obs} = 4$ of the whole MRS stellar source sample (i.e., adding the $\log R_\mathrm{H\alpha}^\mathrm{median} > -4.85$ stellar sources to the $\log R_\mathrm{H\alpha}^\mathrm{median} < -4.85$ stellar sources),
and the relevant result will be discussed at the end of this subsection.

It can be seen from Figures \ref{fig:ratiostdextrha_histogram}a and \ref{fig:ratiostdextrha_histogram}d that the general trend of the $R_{\mathrm{H}{\alpha}}^\mathrm{STD} / R_{\mathrm{H}{\alpha}}^\mathrm{EXT}$ histograms of the MRS subsamples and the normal distribution model is consistent, that is,
the probability density of the smaller $R_{\mathrm{H}{\alpha}}^\mathrm{STD} / R_{\mathrm{H}{\alpha}}^\mathrm{EXT}$ values (left part of the histograms) is generally higher than that of the larger $R_{\mathrm{H}{\alpha}}^\mathrm{STD} / R_{\mathrm{H}{\alpha}}^\mathrm{EXT}$ values (right part of the histograms).
On the other hand, the fine profiles of the histograms do exhibit discrepancies between the $R_{\mathrm{H}{\alpha}}^\mathrm{STD} / R_{\mathrm{H}{\alpha}}^\mathrm{EXT}$ distributions of the MRS sample and the normal distribution model. 
That is, the probability density of the smallest $R_{\mathrm{H}{\alpha}}^\mathrm{STD} / R_{\mathrm{H}{\alpha}}^\mathrm{EXT}$ values of the MRS sample is lower than that of the normal distribution model, 
while the probability density of the medium and large $R_{\mathrm{H}{\alpha}}^\mathrm{STD} / R_{\mathrm{H}{\alpha}}^\mathrm{EXT}$ values of the MRS sample is higher than that of the normal distribution model.
This is true for both the $N_\mathrm{obs} = 3$ and $N_\mathrm{obs} = 4$ subsamples.
The normal distribution model represents random fluctuation of H$\alpha$ activity;
the discrepancies of the $R_{\mathrm{H}{\alpha}}^\mathrm{STD} / R_{\mathrm{H}{\alpha}}^\mathrm{EXT}$ distribution of the MRS sample from that of the normal distribution model mean that at lease part of the stellar sources in the MRS sample have regular variations of H$\alpha$ activity instead of random fluctuations.

In order to find out which part of the stellar sources dominates the discrepancies,
we divide the MRS stellar sources employed in Figures \ref{fig:ratiostdextrha_histogram}a and \ref{fig:ratiostdextrha_histogram}d by the line of $\log R_\mathrm{H\alpha}^\mathrm{EXT} = -5.80$ (see the dashed lines in Figures \ref{fig:ratiostdextrha_histogram}c and \ref{fig:ratiostdextrha_histogram}f). 
The upper envelope of the $\log R_\mathrm{H\alpha}^\mathrm{EXT}$ versus $\log R_\mathrm{H\alpha}^\mathrm{median}$ distribution described in Section \ref{sec:result_logextrha_with_logmedianrha} is roughly above this line. 
In Figures \ref{fig:ratiostdextrha_histogram}b and \ref{fig:ratiostdextrha_histogram}e (middle column of Figure \ref{fig:ratiostdextrha_histogram}), we show the histograms of $R_{\mathrm{H}{\alpha}}^\mathrm{STD} / R_{\mathrm{H}{\alpha}}^\mathrm{EXT}$ for the MRS subsamples above the $\log R_\mathrm{H\alpha}^\mathrm{EXT} = -5.80$ line (i.e., adding screening condition of $\log R_\mathrm{H\alpha}^\mathrm{EXT} > -5.80$).
It can be seen from Figures \ref{fig:ratiostdextrha_histogram}b and~\ref{fig:ratiostdextrha_histogram}e that the discrepancies between the $R_{\mathrm{H}{\alpha}}^\mathrm{STD} / R_{\mathrm{H}{\alpha}}^\mathrm{EXT}$ distributions of the MRS sample and the normal distribution model are enhanced by introducing the $\log R_\mathrm{H\alpha}^\mathrm{EXT} > -5.80$ condition.
This result indicates that the discrepancies are mainly dominated by the stellar sources with large $\log R_\mathrm{H\alpha}^\mathrm{EXT}$ values (upper part of the diagrams in Figures \ref{fig:ratiostdextrha_histogram}c and \ref{fig:ratiostdextrha_histogram}f).
An enhanced deviation of the $R_{\mathrm{H}{\alpha}}^\mathrm{STD} / R_{\mathrm{H}{\alpha}}^\mathrm{EXT}$ distribution of the large-$\log R_\mathrm{H\alpha}^\mathrm{EXT}$ MRS stellar sources with $\log R_\mathrm{H\alpha}^\mathrm{median} < -4.85$ from that of the normal distribution model suggests more regular variations of H$\alpha$ activity of these stellar sources than the stellar sources with smaller $\log R_\mathrm{H\alpha}^\mathrm{EXT}$.

If we add the $\log R_\mathrm{H\alpha}^\mathrm{median} > -4.85$ stellar sources to the $\log R_\mathrm{H\alpha}^\mathrm{median} < -4.85$ stellar sources (see 
Appendix \ref{sec:appendix_histratiostdextrha_wholesample} and Figure~\ref{fig:ratiostdextrha_histogram_wholesample}),
the discrepancies between the $R_{\mathrm{H}{\alpha}}^\mathrm{STD} / R_{\mathrm{H}{\alpha}}^\mathrm{EXT}$ histograms of the MRS sample and the normal distribution model are reduced,
which suggests that the H$\alpha$ variations of the $\log R_\mathrm{H\alpha}^\mathrm{median} > -4.85$ stellar sources are more likely to be random fluctuations.

\subsection{Re-examine the bottom envelope of the $\log R_{\mathrm{H}{\alpha}}^\mathrm{EXT}$ versus $\log R_\mathrm{H\alpha}^\mathrm{median}$ distribution} \label{sec:discu_bottom_envelope}

In the analysis in Section \ref{sec:result_logextrha_with_logmedianrha}, we focus on the top envelope of the $\log R_{\mathrm{H}{\alpha}}^\mathrm{EXT}$ versus $\log R_\mathrm{H\alpha}^\mathrm{median}$ distribution, and the lower envelope of the distribution is mostly a truncation effect of the $R_{\mathrm{H}{\alpha}}^\mathrm{EXT} > 3 \times \delta R_{\mathrm{H}{\alpha}}^\mathrm{EXT}$ screening condition.
In this subsection, we examine whether there exists a physical bottom envelope of the $\log R_{\mathrm{H}{\alpha}}^\mathrm{EXT}$ versus $R_{\mathrm{H}{\alpha}}^\mathrm{median}$ distribution if the time span of observations ($T_\mathrm{span}$) is sufficiently long and the number of observations ($N_\mathrm{obs}$) is sufficiently large. 

\begin{figure}
  \centering
  \includegraphics[width=0.56\textwidth]{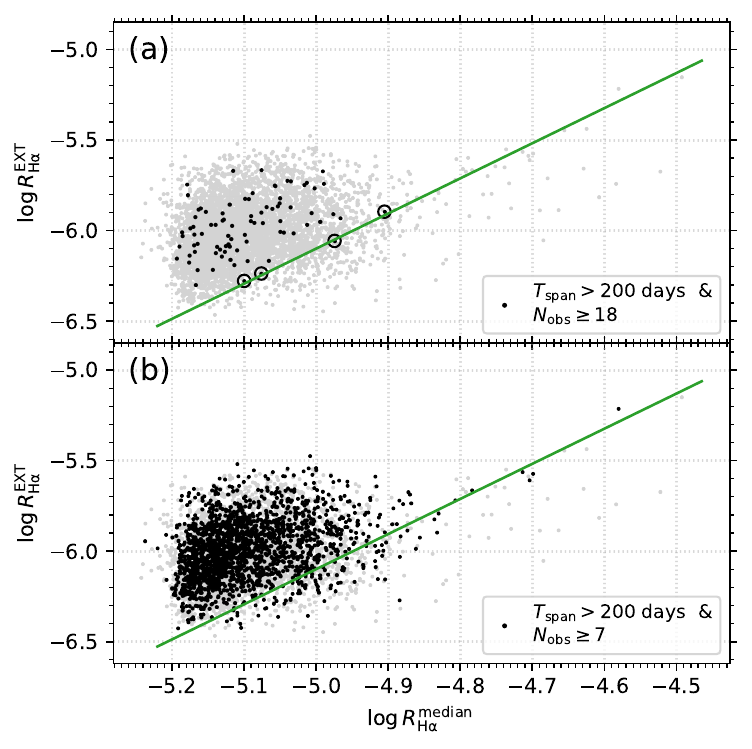}
  \caption{Scatter plots of $\log R_\mathrm{H\alpha}^\mathrm{EXT}$ versus $\log R_\mathrm{H\alpha}^\mathrm{median}$ (data points in black) for the subsamples of stellar sources with (a) $T_\mathrm{span} > 200$ days \& $N_\mathrm{obs} \geq 18$ and (b) $T_\mathrm{span} > 200$ days \& $N_\mathrm{obs} \geq 7$. 
   The whole stellar source sample (as employed in Figure \ref{fig:logextrha_vs_logmedianrha_density}) is displayed in light gray in the background for reference.
   The solid green line in the two panels is the linear fit of the bottom envelope shown in panel (a) (described by Equation~(\ref{equ:logextrha_vs_logmedianrha_bottom_envelope})).
   The circle symbols in panel (a) mark the data points used for the fit.
   }
  \label{fig:logextrha_vs_logmedianrha_bottom_envelope}
\end{figure}

In Figure \ref{fig:logextrha_vs_logmedianrha_bottom_envelope}a, we show the scatter plot of $\log R_\mathrm{H\alpha}^\mathrm{EXT}$ versus $\log R_\mathrm{H\alpha}^\mathrm{median}$ for the subsample of stellar sources with $T_\mathrm{span} > 200$ days \& $N_\mathrm{obs} \geq 18$ (data points in black; $84$ in total). 
The $T_\mathrm{span} > 200$ days screening condition ensures a long time span of observation.
The $N_\mathrm{obs} \geq 18$ screening condition ensures a large number of observation,
and the stellar source subsample subject to this condition is less affected by the truncation effect of $R_{\mathrm{H}{\alpha}}^\mathrm{EXT} > 3 \times \delta R_{\mathrm{H}{\alpha}}^\mathrm{EXT}$, as demonstrated in Figure \ref{fig:extrha_histogram_and_vs_numobs}b.
The whole stellar source sample (as employed in Figure \ref{fig:logextrha_vs_logmedianrha_density}) is displayed in light gray in the background in Figure \ref{fig:logextrha_vs_logmedianrha_bottom_envelope}a for reference. 

The scatter plot in Figure \ref{fig:logextrha_vs_logmedianrha_bottom_envelope}a shows a clue to the physical bottom envelope of the $\log R_{\mathrm{H}{\alpha}}^\mathrm{EXT}$ versus $\log R_\mathrm{H\alpha}^\mathrm{median}$ distribution.
We perform a linear fit of the bottom envelope exhibited by the data points (in black) in Figure \ref{fig:logextrha_vs_logmedianrha_bottom_envelope}a, 
and the data points located on the bottom envelope used for the fit are marked with circle symbols.
The equation of the fitted bottom envelope is
\begin{equation} \label{equ:logextrha_vs_logmedianrha_bottom_envelope}
  \log R_{\mathrm{H}{\alpha}}^\mathrm{EXT} = 1.94 \times \log R_\mathrm{H\alpha}^\mathrm{median} + 3.60,
\end{equation}
which is displayed as a solid green line in Figure \ref{fig:logextrha_vs_logmedianrha_bottom_envelope}a.
The coefficient of the $\log R_\mathrm{H\alpha}^\mathrm{median}$ term in Equation (\ref{equ:logextrha_vs_logmedianrha_bottom_envelope}) is $1.94$, which is approximate to $2$, indicating that along the bottom envelope $R_\mathrm{H\alpha}^\mathrm{EXT}$ is roughly proportional to the square of $R_\mathrm{H\alpha}^\mathrm{median}$.

The stellar source subsample used in Figure \ref{fig:logextrha_vs_logmedianrha_bottom_envelope}a does not cover the data points with $\log R_\mathrm{H\alpha}^\mathrm{median} > -4.85$ (higher activity intensity) in the right part of the diagram.
As a comparison, in Figure \ref{fig:logextrha_vs_logmedianrha_bottom_envelope}b we show the scatter plot of $\log R_\mathrm{H\alpha}^\mathrm{EXT}$ versus $\log R_\mathrm{H\alpha}^\mathrm{median}$ for the subsample of stellar sources with a looser condition of $T_\mathrm{span} > 200$ days \& $N_\mathrm{obs} \geq 7$ (data points in black; 1,769 in total) as well as the fitted line described by Equation (\ref{equ:logextrha_vs_logmedianrha_bottom_envelope}).
The data points in Figure \ref{fig:logextrha_vs_logmedianrha_bottom_envelope}b are much denser,
and the fitted line coincides with the bottom envelope of the dense area of the data points.
In addition, it can be seen from Figure~\ref{fig:logextrha_vs_logmedianrha_bottom_envelope}b that the fitted line also coincides with the distribution of the data points with $\log R_\mathrm{H\alpha}^\mathrm{median} > -4.85$ in the right part of the diagram,
and those data points with $\log R_\mathrm{H\alpha}^\mathrm{median} > -4.85$ (in black) are distributed roughly along the top envelope of the $\log R_\mathrm{H\alpha}^\mathrm{EXT}$ versus $\log R_\mathrm{H\alpha}^\mathrm{median}$ distribution of the whole sample (shown in light gray).
This result implies that the physical origin of the bottom envelope in the left part of the diagram ($\log R_\mathrm{H\alpha}^\mathrm{median} < -4.85$; with lower activity intensity) and the top envelope in the right part of the diagram ($\log R_\mathrm{H\alpha}^\mathrm{median}>-4.85$; with higher activity intensity) might be the same, that is,
the intrinsic random fluctuation of the H$\alpha$ activity that is only related to the stellar activity intensity.
Since along the fitted line $R_\mathrm{H\alpha}^\mathrm{EXT}$ is roughly proportional to the square of $R_\mathrm{H\alpha}^\mathrm{median}$ (see Equation (\ref{equ:logextrha_vs_logmedianrha_bottom_envelope})),
for the lower $R_\mathrm{H\alpha}^\mathrm{median}$ values (in the left part of the diagram), this intrinsic fluctuation is not prominent relative to other factors that can lead to H$\alpha$ variability (such as rotational modulation and stellar activity cycles);
but for the higher $R_\mathrm{H\alpha}^\mathrm{median}$ values (in the right part of the diagram), the intrinsic fluctuation dominates the H$\alpha$ variability and can lead to very large $R_\mathrm{H\alpha}^\mathrm{EXT}$ values.

\subsection{Physical implications of the H$\alpha$ variability distribution} \label{sec:discu_physical_implications}

In the previous analyses, we have investigated various aspects of the distribution of H$\alpha$ variability (represented by $\log R_{\mathrm{H}{\alpha}}^\mathrm{EXT}$) for solar-like stars.
In this subsection, we discuss physical implications of the H$\alpha$ variability distribution.

1. Stratification of stellar H$\alpha$ activity levels.

The stratification of stellar H$\alpha$ activity levels refers to the distinct H$\alpha$ activity properties for different activity levels of solar-like stars. 
It can be manifested by the distributions of the H$\alpha$ activity indices with stellar atmospheric parameters (\citealt{2023Ap&SS.368...63H, 2024NatSR..1417962H}), and by the correlations between the activity indices of different Hydrogen Balmer lines (\citealt{2024NatSR..1417962H}).
In this study, the stratification of stellar H$\alpha$ activity levels is manifested by the distribution of H$\alpha$ variability ($\log R_{\mathrm{H}{\alpha}}^\mathrm{EXT}$) with H$\alpha$ activity intensity ($\log R_\mathrm{H\alpha}^\mathrm{median}$). 
As the result obtained in Section \ref{sec:result_logextrha_with_logmedianrha}, the distribution of $\log R_{\mathrm{H}{\alpha}}^\mathrm{EXT}$ of the stellar sources with higher activity levels ($\log R_\mathrm{H\alpha}^\mathrm{median} > -4.85$) is distinctly different from that of the stellar sources with lower activity levels ($\log R_\mathrm{H\alpha}^\mathrm{median} < -4.85$).
For the stellar source category with higher activity levels, the top envelope of the $\log R_{\mathrm{H}{\alpha}}^\mathrm{EXT}$ versus $\log R_\mathrm{H\alpha}^\mathrm{median}$ distribution is largely along a positive correlation line and shows a gradual increase with $\log R_\mathrm{H\alpha}^\mathrm{median}$;
while for the stellar source category with lower activity levels, the top envelope of the distribution first increases and then decreases with $\log R_\mathrm{H\alpha}^\mathrm{median}$, 
and finally intersects the top envelope of the stellar source category with higher activity levels at about $\log R_\mathrm{H\alpha}^\mathrm{median} = -4.85$.
The result in Section \ref{sec:result_logextrha_in_teff_logmedianrha_space} also shows the different properties of the $\log R_{\mathrm{H}{\alpha}}^\mathrm{EXT}$ distribution in the $T_\mathrm{eff}$ -- $\log R_{\mathrm{H}{\alpha}}^\mathrm{median}$ parameter space for the stellar source categories with lower and higher activity levels.
Because of their distinct behaviors of H$\alpha$ variability,
in our analysis the two categories of stellar sources are separated where appropriate, as performed in the previous sections.

2. Two types of H$\alpha$ variability: long-term cyclic and random.

The analysis of the $\log R_{\mathrm{H}{\alpha}}^\mathrm{EXT}$ distribution with $T_\mathrm{span}$ in Section \ref{sec:result_logextrha_with_tspan} shows that,
for the stellar source category with lower activity levels ($\log R_\mathrm{H\alpha}^\mathrm{median} < -4.85$),
the large $\log R_{\mathrm{H}{\alpha}}^\mathrm{EXT}$ values near the top envelope of the $\log R_{\mathrm{H}{\alpha}}^\mathrm{EXT}$ versus $\log R_\mathrm{H\alpha}^\mathrm{median}$ distribution tend to be associated with the long-term variations of H$\alpha$ activity.
The discussion in Section \ref{sec:discu_extrha_stdrha} further shows that this large-$\log R_{\mathrm{H}{\alpha}}^\mathrm{EXT}$ subsample also tends to have more regular variations of H$\alpha$ activity.
The long-term and regular variations of H$\alpha$ activity suggest that the H$\alpha$ variability of the subsample (with lower activity levels and large $\log R_{\mathrm{H}{\alpha}}^\mathrm{EXT}$) might be associated with stellar activity cycles,
i.e., long-term cyclic variations of H$\alpha$ activity.

For the stellar source category with higher activity levels ($\log R_\mathrm{H\alpha}^\mathrm{median} > -4.85$),
the analysis in Section \ref{sec:result_logextrha_with_tspan} shows that there is no evidence of distinct distribution configurations between the long-term and short-term variations of H$\alpha$ activity.
The discussion in Section \ref{sec:discu_extrha_stdrha} and Appendix \ref{sec:appendix_histratiostdextrha_wholesample} further shows that the H$\alpha$ variability of the $\log R_\mathrm{H\alpha}^\mathrm{median} > -4.85$ category is more likely to be associated with random fluctuations.
In addition, the discussion in Section \ref{sec:discu_bottom_envelope} shows that the bottom envelope of the $\log R_{\mathrm{H}{\alpha}}^\mathrm{EXT}$ versus $\log R_\mathrm{H\alpha}^\mathrm{median}$ distribution of the $\log R_\mathrm{H\alpha}^\mathrm{median} < -4.85$ stellar source category coincides with the top envelope of the $\log R_\mathrm{H\alpha}^\mathrm{median} > -4.85$ category (with $R_{\mathrm{H}{\alpha}}^\mathrm{EXT}$ being roughly proportional to the square of $R_\mathrm{H\alpha}^\mathrm{median}$),
and their physical origins may all come from the intrinsic random fluctuation of H$\alpha$ activity that is only related to the stellar activity intensity.
This basal random variation of H$\alpha$ activity is relatively weak for the stellar sources with lower activity levels ($\log R_\mathrm{H\alpha}^\mathrm{median} < -4.85$) and constitutes the bottom envelope of the $\log R_{\mathrm{H}{\alpha}}^\mathrm{EXT}$ distribution,
but for the stellar sources with higher activity levels ($\log R_\mathrm{H\alpha}^\mathrm{median} > -4.85$) it is prominent and plays dominant role in the H$\alpha$ variability.

The observation projects dedicated to the long-term monitoring of stellar activities had revealed that the stellar activity cycles with long-term and regular variations of stellar activity are more often observed for the stellar objects with low activity levels, 
and young stars with high levels of activity rarely display a long-term cyclic variation (e.g., \citealt{1995ApJ...438..269B}), 
which is consistent with the above results and discussions.

\section{Conclusion} \label{sec:con}

In this work, we performed a systematic investigation on the variability of H$\alpha$ chromospheric activity of solar-like stars by using the MRS time-domain data of LAMOST for the first time.
We select MRS time-domain sample of solar-like stars from the LAMOST DR10 v1.0 which encompasses the data from the first five years of MRS.
Strict screening conditions are employed to ensure the quality of the selected MRS spectra and the consistency between the H$\alpha$ spectra of multiple observations of each stellar source.
In the selected MRS time-domain sample employed in this work, 
the number of MRS observations of each stellar source ($N_\mathrm{obs}$) ranges from 3 to 31,
and the time span of multiple observations of each stellar source ($T_\mathrm{span}$) ranges from several days to several years.
We use $R_\mathrm{H\alpha}$ index to measure the H$\alpha$ activity intensity of a spectrum,
and utilize the median of the $R_\mathrm{H\alpha}$ values of multiple observations ($R_\mathrm{H\alpha}^\mathrm{median}$) as the representative H$\alpha$ activity intensity of a stellar source.  
The H$\alpha$ variability of a stellar source is indicated by the extent of $R_\mathrm{H\alpha}$ fluctuation ($R_\mathrm{H\alpha}^\mathrm{EXT}$) of multiple observations.
The main analysis results are as follows.

1. Our sample shows that the peak of the $R_\mathrm{H\alpha}^\mathrm{EXT}$ distribution is at about $10^{-6}$,
which is one order of magnitude smaller than the values of $R_\mathrm{H\alpha}$ (about $10^{-5}$).

2. The distribution of $\log R_\mathrm{H\alpha}^\mathrm{EXT}$ versus $\log R_\mathrm{H\alpha}^\mathrm{median}$ reveals the distinct behaviors between the stellar source categories with lower ($\log R_\mathrm{H\alpha}^\mathrm{median} < -4.85$) and higher ($\log R_\mathrm{H\alpha}^\mathrm{median} > -4.85$) activity intensity.
For the former stellar source category, the top envelope of the distribution first increases with $\log R_{\mathrm{H}{\alpha}}^\mathrm{median}$,
reaching $R_{\mathrm{H}{\alpha}}^\mathrm{EXT} = 0.35 \times R_{\mathrm{H}{\alpha}}^\mathrm{median}$ around $\log R_{\mathrm{H}{\alpha}}^\mathrm{median} \sim -5.1$,
and then decreases with the increase of $\log R_{\mathrm{H}{\alpha}}^\mathrm{median}$ from about $\log R_{\mathrm{H}{\alpha}}^\mathrm{median} \sim -5.0$;
while for the latter category, the top envelope of the distribution is largely along a positive correlation line and shows a gradual increase with $\log R_\mathrm{H\alpha}^\mathrm{median}$.
The distinction between the $\log R_\mathrm{H\alpha}^\mathrm{EXT}$ versus $\log R_\mathrm{H\alpha}^\mathrm{median}$ distributions of the stellar source categories with higher and lower activity intensity supports the concept of stratification of stellar H$\alpha$ activity levels.

3. The distribution of $\log R_\mathrm{H\alpha}^\mathrm{EXT}$ in the $T_\mathrm{eff}$ -- $\log R_\mathrm{H\alpha}^\mathrm{median}$ parameter space reveals that,
for the $\log R_\mathrm{H\alpha}^\mathrm{median} < -4.85$ stellar source category,
the peak of the $\log R_\mathrm{H\alpha}^\mathrm{EXT}$ values appears around $T_\mathrm{eff} \sim 5600$\,K;
while for the $\log R_\mathrm{H\alpha}^\mathrm{median} > -4.85$ stellar source category,
the largest $\log R_\mathrm{H\alpha}^\mathrm{EXT}$ values correspond to the smallest $T_\mathrm{eff}$ in our sample.

4. The distribution of $\log R_\mathrm{H\alpha}^\mathrm{EXT}$ for different $T_\mathrm{span}$ scales reveals that,
for the $\log R_\mathrm{H\alpha}^\mathrm{median} < -4.85$ stellar source category,
the large $\log R_\mathrm{H\alpha}^\mathrm{EXT}$ values near the top envelope of the $\log R_\mathrm{H\alpha}^\mathrm{EXT}$ versus $\log R_\mathrm{H\alpha}^\mathrm{median}$ distribution tend to be associated with the long-term ($T_\mathrm{span} > 200$~days) variations of H$\alpha$ activity,
and the smaller $\log R_\mathrm{H\alpha}^\mathrm{EXT}$ values tend to be associated with the short-term ($T_\mathrm{span} < 200$~days) variations of H$\alpha$ activity;
while for the $\log R_\mathrm{H\alpha}^\mathrm{median} > -4.85$ stellar source category,
this distinction of $\log R_\mathrm{H\alpha}^\mathrm{EXT}$ distribution for different $T_\mathrm{span}$ scales is not exhibited.

5. The comparison of the $R_\mathrm{H\alpha}^\mathrm{STD} / R_\mathrm{H\alpha}^\mathrm{EXT}$ distribution of the MRS time-domain sample with that of the normal distribution model shows that,
for the $\log R_\mathrm{H\alpha}^\mathrm{median} < -4.85$ stellar source category,
the large-$\log R_\mathrm{H\alpha}^\mathrm{EXT}$ objects near the top envelope of the $\log R_\mathrm{H\alpha}^\mathrm{EXT}$ versus $\log R_\mathrm{H\alpha}^\mathrm{median}$ distribution tend to have more regular variations of stellar activity (might be associated with stellar activity cycles);
while the H$\alpha$ variations of the $\log R_\mathrm{H\alpha}^\mathrm{median} > -4.85$ category are more likely to be random fluctuations. 

The results obtained in this work provide an overview of the stellar H$\alpha$ variability properties of solar-like stars, which can be verified and enriched with more time-domain observation data of LAMOST-MRS in the future.
The methods developed in this work can be helpful for the future studies of large-sample statistics of stellar activity variability based on the sky survey projects as well as the studies on stellar variability of individual stars.

\vspace{10mm}

\begin{appendices}

\renewcommand\thefigure{\arabic{figure}}
\setcounter{figure}{13}
\renewcommand\thetable{\arabic{table}}
\setcounter{table}{1}

\section{Distribution diagrams of the $I_{\mathrm{H}{\alpha}}$ index for the MRS time-domain sample of solar-like stars employed in this work} \label{sec:appendix_iha}

\begin{figure*}
  \centering
  \includegraphics[width=0.95\textwidth]{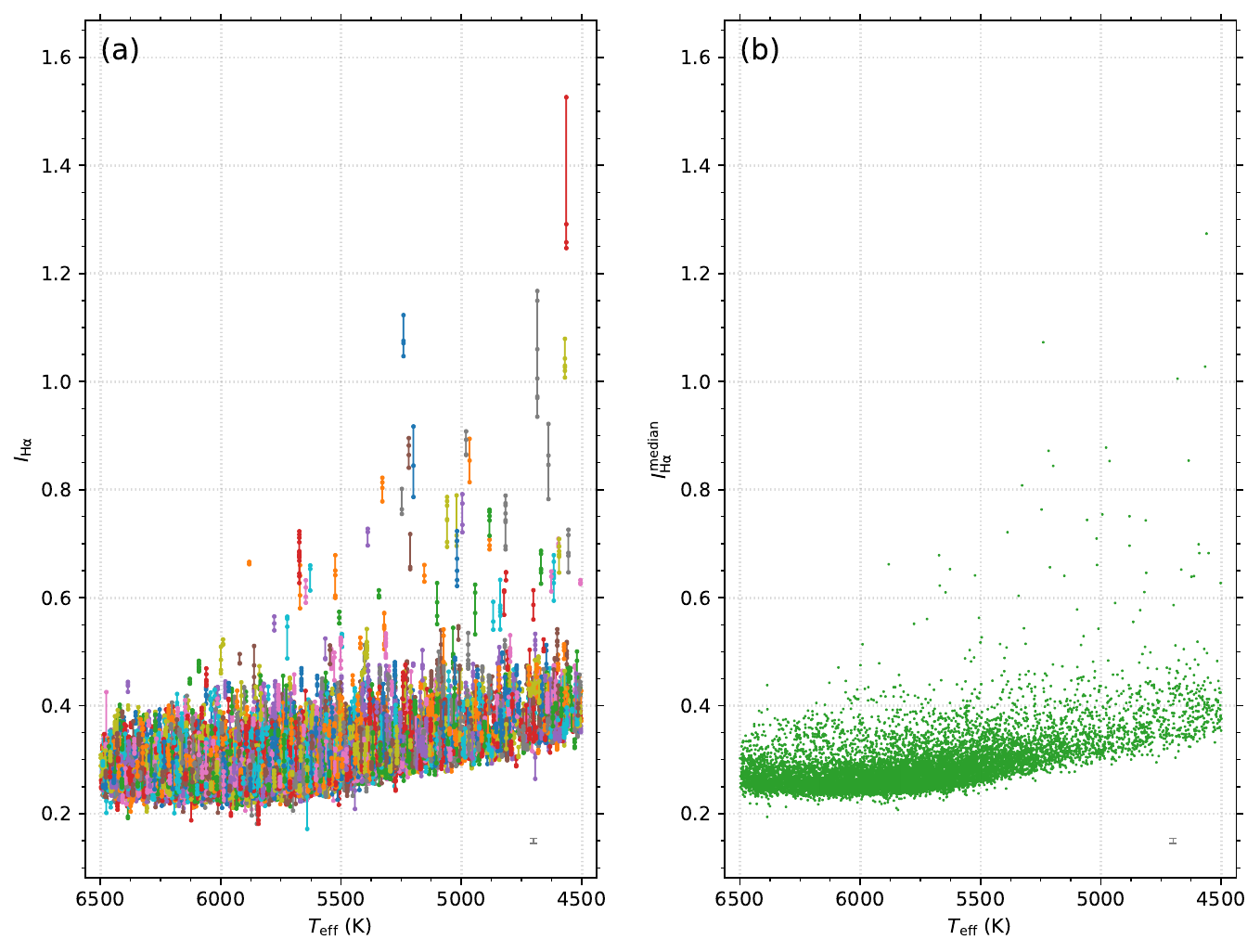}
  \caption{Distribution of $I_\mathrm{H\alpha}$ index with stellar effective temperature for (a) the spectra and (b) the stellar sources (using $I_\mathrm{H\alpha}^\mathrm{median}$) in the MRS time-domain sample of solar-like stars employed in this work.
In panel (a), the $I_\mathrm{H\alpha}$ values from the multiple observations of the same stellar source are connected by a vertical line segment,
and the different line segments (stellar sources) are displayed in a variety of colors for ease of distinction.
The error bar ($\pm \delta I_{\mathrm{H}{\alpha}}$) in the bottom right corner of each panel shows the median of the uncertainties of the $I_\mathrm{H\alpha}$ values (about $0.005$).
  }
  \label{fig:iha_vs_teff_mrs_timedomain_spectra}
\end{figure*}

$I_{\mathrm{H}{\alpha}}$ index is the observed mean flux in the central band of the H$\alpha$ line normalized by the continuum flux.
Its relation to the $R_{\mathrm{H}{\alpha}}$ index can be described by a multiplier factor, as shown in Equation (\ref{equ:Rindex}),
and the factor in turn is a function of stellar atmospheric parameters.
Since $I_{\mathrm{H}{\alpha}}$ is a quantity directed derived from the observed MRS spectral data and does not depend on a specific stellar atmospheric model,
its distribution may be helpful for certain research topics.
In Figure \ref{fig:iha_vs_teff_mrs_timedomain_spectra}, we show the distribution of the $I_{\mathrm{H}{\alpha}}$ index values with stellar effective temperature for the MRS time-domain sample of solar-like stars employed in this work.

The layout of Figure \ref{fig:iha_vs_teff_mrs_timedomain_spectra} is similar to the distribution diagrams of the $R_{\mathrm{H}{\alpha}}$ index shown in Figure \ref{fig:rha_vs_teff_mrs_timedomain_spectra}.
Figure \ref{fig:iha_vs_teff_mrs_timedomain_spectra}a displays the $I_{\mathrm{H}{\alpha}}$ distribution for the spectra in the MRS time-domain sample of solar-like stars,
and Figure \ref{fig:iha_vs_teff_mrs_timedomain_spectra}b displays the  $I_{\mathrm{H}{\alpha}}$ distribution for the stellar sources using $I_\mathrm{H\alpha}^\mathrm{median} = \mathrm{median}(\{I_\mathrm{H\alpha}^i, i=1, \cdots, N_\mathrm{obs}\})$,
where $\{I_\mathrm{H\alpha}^i\}$ is the set of multiple $I_{\mathrm{H}{\alpha}}$ measurements of each source.
The median of the uncertainties of the $I_{\mathrm{H}{\alpha}}$ values ($\delta I_{\mathrm{H}{\alpha}}$) is about $0.005$, which is shown as an error bar ($\pm \delta I_{\mathrm{H}{\alpha}}$) in the bottom right corner of the diagrams in Figure~\ref{fig:iha_vs_teff_mrs_timedomain_spectra}.

\section{Criterion for identifying the MRS spectra contaminated by the H\,II regions} \label{sec:appendix_HII_regions} 

As described in Section \ref{sec:sample_spcond_HII}, the MRS spectra contaminated by the H\,II regions can be identified using the N\,II spectral lines on the two sides of the H$\alpha$ line.
The two N\,II lines are characteristic lines of H\,II regions, and are very weak for solar-like stars.
By recognizing the prominent N\,II lines, the MRS spectra contaminated by the H\,II regions can be identified.

\begin{figure}
  \centering
  \includegraphics[width=0.49\textwidth]{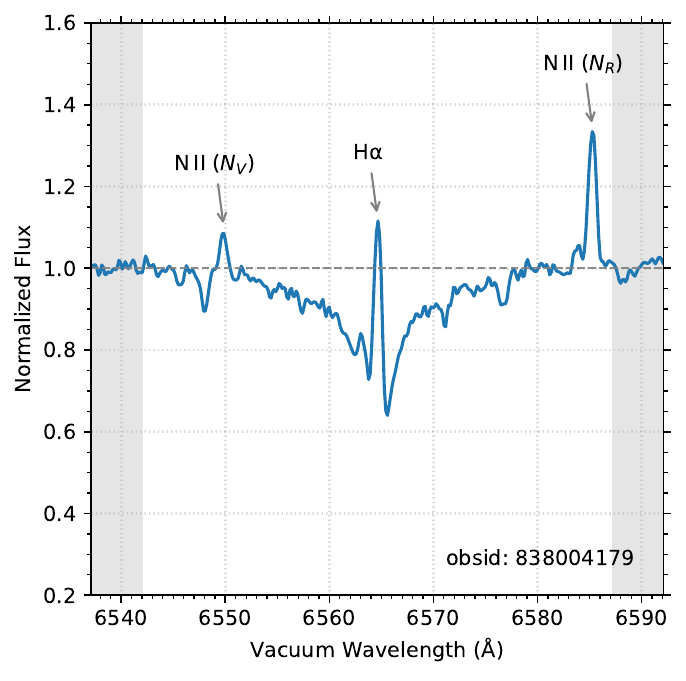}
  \caption{Example MRS spectrum (obsid: 838004179) of the N\,II lines from H\,II regions.
  The spectral flux is normalized,
  and the wavelength shift caused by radial velocity is corrected.
  The two N\,II lines (for deriving the $N_V$ and $N_R$ indices, respectively) as well as the H$\alpha$ line are labeled.
  The two shaded areas indicate the two continuum bands used for evaluating the $N_V$ and $N_R$ indices.
  The horizontal dashed line represents the normalized continuum of the spectrum.} 
  \label{fig:nvnr_example_spectrum}
\end{figure}

\begin{figure*}
  \centering
  \includegraphics[width=0.97\textwidth]{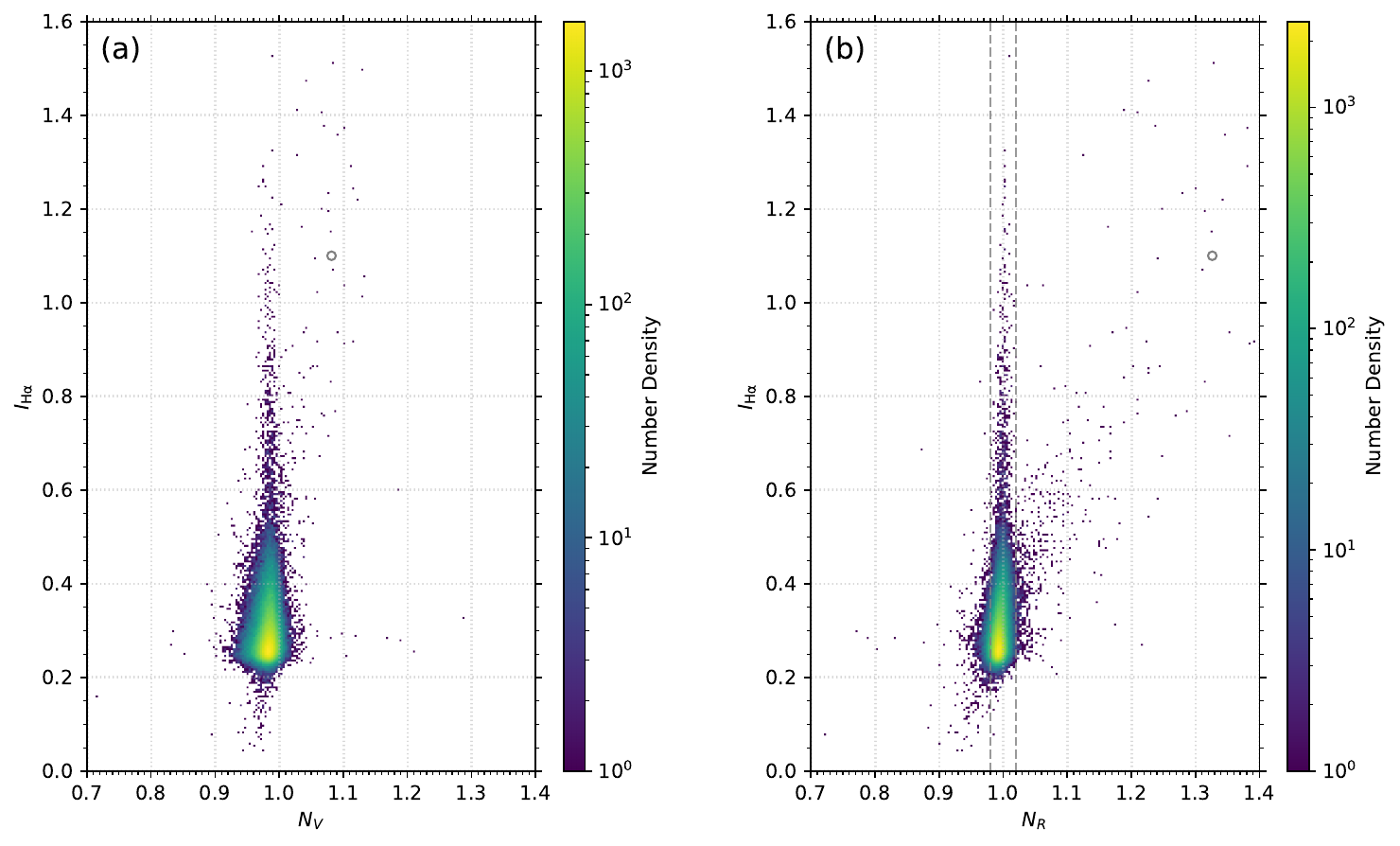}
  \caption{
    Diagrams illustrating the relations of (a) $N_V$ index and (b) $N_R$ index derived from the N\,II lines with $I_\mathrm{H\alpha}$ index derived from the H$\alpha$ line for the sample of MRS spectra after the screening conditions described in Sections \ref{sec:sample_spcond_solarlike}--\ref{sec:sample_spcond_rotvel}.
    Color scale indicates number density.
    The two vertical dashed lines in panel (b) represent $N_R = 0.98$ and $1.02$;
    the data points outside the region enclosed by the two lines are considered to be contaminated by the H\,II regions (see main text for details).
    The circle symbol in both panels indicates the position of the example spectrum of H\,II regions employed in Figure \ref{fig:nvnr_example_spectrum}. 
  }
  \label{fig:iha_vs_nvnr}
\end{figure*}

Figure \ref{fig:nvnr_example_spectrum} shows an example MRS spectrum of the N\,II lines from H\,II regions, in which the H$\alpha$ emission line of H\,II regions is located in between the two N\,II lines.
The vacuum wavelengths of the two N\,II line are $6549.86$\,{\AA} and $6585.27$\,{\AA}, respectively \citep{2002AJ....123..485S}.
We calculate the mean fluxes in the central bands of the N\,II lines ($6549.860 \pm 0.125$\,{\AA} and $6585.270 \pm 0.125$\,{\AA}, respectively, with band width being 0.25\,{\AA}), 
and then normalize the central fluxes by the fluxes in two nearby continuum bands.
The two continuum bands (see illustration in Figure \ref{fig:nvnr_example_spectrum}) are the same as used for evaluating the $I_\mathrm{H\alpha}$ index (see Section \ref{sec:measure_activityintensity}).
Since the two continuum bands are asymmetrical relative to the N\,II lines,
we adopt the flux normalization scheme proposed in \citet{2024NatSR..1417962H},
in which the weight of the flux of a continuum band is inversely proportional to the distance of the continuum band from the spectral line.
The results of the normalized central fluxes of the two N\,II lines are named $N_V$ index and $N_R$ index for the two N\,II lines on the violet side and the red side of the H$\alpha$ line, respectively. 
We investigate the relations of the $N_V$ and $N_R$ indices obtained from the N\,II lines with the $I_\mathrm{H\alpha}$ index obtained from the H$\alpha$ line for the sample of MRS spectra after the screening conditions described in Sections \ref{sec:sample_spcond_solarlike}--\ref{sec:sample_spcond_rotvel};
the result is shown in Figure \ref{fig:iha_vs_nvnr}.
Panel (a) of Figure \ref{fig:iha_vs_nvnr} is for the relation of $N_V$ with $I_\mathrm{H\alpha}$ and panel (b) is for the relation of $N_R$ with $I_\mathrm{H\alpha}$.
The position of the example spectrum of H\,II regions employed in Figure \ref{fig:nvnr_example_spectrum} is indicated with a circle symbol in the two panels of Figure \ref{fig:iha_vs_nvnr}.

It can be seen from Figure \ref{fig:iha_vs_nvnr} that the distribution of $I_\mathrm{H\alpha}$ index with $N_V$ (or $N_R$) index exhibits two branches, the diagonal branch and the vertical branch.
In the diagonal branch, there is a positive correlation trend between the $I_\mathrm{H\alpha}$ and $N_V$ (or $N_R$) values,
and the example spectrum of H\,II regions employed in Figure \ref{fig:nvnr_example_spectrum} is located in this branch;
while in the vertical branch, $I_\mathrm{H\alpha}$ does not show correlation with $N_V$ (or $N_R$).
Therefore, the diagonal branch is the sample of MRS spectra contaminated with the emissions from H\,II regions,
and the vertical branch is the sample directly from stars.
By comparing the diagrams of $N_V$ (Figure \ref{fig:iha_vs_nvnr}a) and $N_R$ (Figure \ref{fig:iha_vs_nvnr}b), 
it can be found that the separation of the two branches is more prominent for the $N_R$ index;
so the $N_R$ index is a better indicator to distinguish the two branches of samples than the $N_V$ index.
In this work, we utilize the $N_R$ index to identify the MRS spectra contaminated by the H\,II regions.
By carefully examining the diagram of the $N_R$ index in Figure \ref{fig:iha_vs_nvnr}b,
the criterion for the sample in the diagonal branch is determined to be $N_R \le 0.98$ or $N_R \ge 1.02$.
The thresholds of the criterion, $N_R = 0.98$ and $1.02$, are indicated by two vertical dashed lines in Figure \ref{fig:iha_vs_nvnr}b.
The MRS spectra that meet the criterion are considered to be contaminated by the H\,II regions and are removed from our sample.

\section{Criterion for consistency of H$\alpha$ profiles} \label{sec:appendix_profileconsistency}

\subsection{Consistency of H$\alpha$ profiles in the line wings} \label{sec:appendix_profileconsistency_linewings}

\begin{figure*}
  \centering
  \includegraphics[width=0.77\textwidth]{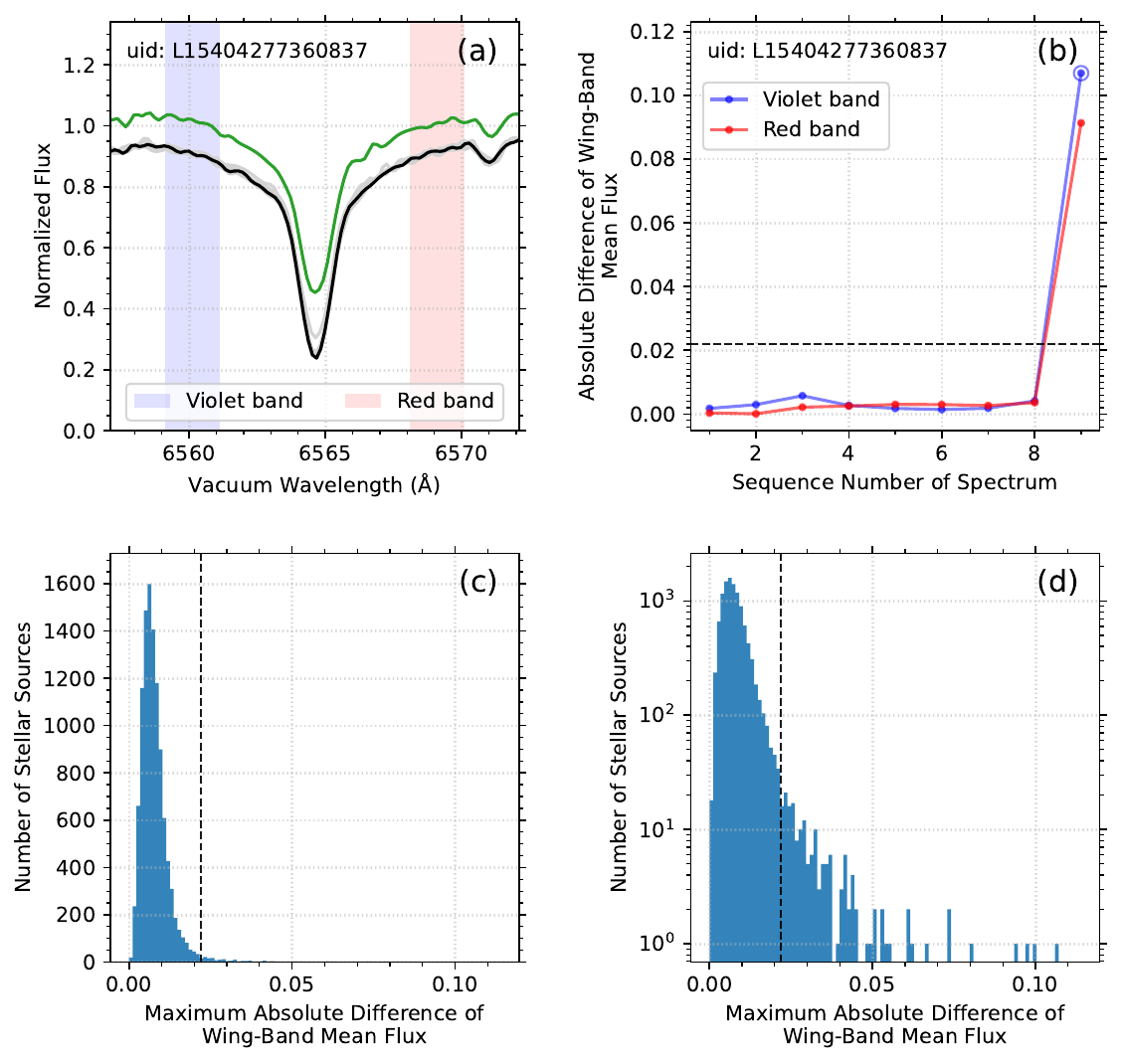}
  \caption{(a) Example MRS H$\alpha$ spectra of a stellar source (uid: L15404277360837; ten spectra in total) that show inconsistent profiles in the line wings.
  The spectral flux is normalized,
  and the wavelength shift caused by radial velocity is corrected.
  The spectra are sorted by their $I_\mathrm{H\alpha}$ index values.
  The spectrum with the smallest $I_\mathrm{H\alpha}$ index value (No. 0; bottom spectrum) is plotted in black,
  and the spectrum with the largest $I_\mathrm{H\alpha}$ index value (No. 9; top spectrum) is plotted in green.
  Other spectra (No. 1 to No. 8) are plotted in light gray.
  The two 2\,{\AA}-wide line-wing bands on the violet side (violet band) and red side (red band) of the H$\alpha$ line for testing the line-wing consistency (with central wavelengths being $\pm 4.5$\,{\AA} from the line center) are indicated by two shaded areas in blue and red, respectively.
  (b) Variation of the absolute differences of the wing-band mean fluxes with the sequence numbers (sorted by $I_\mathrm{H\alpha}$ index) of the spectra.
  The spectrum No. 0 is used as the reference spectrum relative to which the differences of the wing-band mean fluxes of other spectra are calculated,
  and hence does not appear in the plots.
  The curves for the violet and red bands are plotted in blue and red, respectively.
  The circle symbol in the plot indicates the maximum absolute difference value of the wing-band mean fluxes.
  (c) Distribution histogram of the maximum absolute difference values of the wing-band mean fluxes for the stellar source sample after the screening conditions described in Sections \ref{sec:sample_srccond_timedomain} and \ref{sec:sample_srccond_paraconsistency},
  with a linear scale on the vertical axis.
  (d) Same as the histogram in panel (c), but with a logarithmic scale on the vertical axis.
  The horizontal dashed line in panel (b) and the vertical dashed lines in panels (c) and (d) indicate the upper threshold ($0.022$) of the maximum absolute difference of wing-band mean flux adopted in this work for recognizing the consistent H$\alpha$ profiles in the line wings.
  } 
  \label{fig:line_wings_consistency}
\end{figure*}

As described in Section \ref{sec:sample_srccond_profileconsistency},
in order to recognize the stellar sources with consistent H$\alpha$ profiles in the line wings,
we calculate the means of the normalized spectral fluxes in two 2\,{\AA}-wide line-wing bands on the two sides of the H$\alpha$ line (central wavelengths being $\pm 4.5$\,{\AA} from the line center) for each of the MRS time-domain spectra of a stellar source.
Figure \ref{fig:line_wings_consistency}a gives an example of the MRS H$\alpha$ spectra of a stellar source (uid: L15404277360837; ten spectra in total) that show inconsistent profiles in the line wings.
The two line-wing bands on the violet side (violet band) and red side (red band) of the H$\alpha$ line are indicated by two shaded areas in blue and red, respectively, in Figure \ref{fig:line_wings_consistency}a.
The spectra of the stellar source are sorted by their $I_\mathrm{H\alpha}$ index values.
The spectrum with the smallest $I_\mathrm{H\alpha}$ index value (No. 0; bottom spectrum) is plotted in black,
and the spectrum with the largest $I_\mathrm{H\alpha}$ index value (No. 9; top spectrum) is plotted in green.
Other spectra (No. 1 to No. 8) are plotted in light gray.
We calculate the absolute differences of the wing-band mean fluxes between the H$\alpha$ spectrum with the smallest $I_\mathrm{H\alpha}$ index value and other spectra with larger $I_\mathrm{H\alpha}$ index values. 
The variation of the absolute differences of the wing-band mean fluxes with the sequence numbers of the spectra (sorted by $I_\mathrm{H\alpha}$ index) is shown in Figure \ref{fig:line_wings_consistency}b,
and the curves for the violet and red bands are plotted in blue and red, respectively. 
Because the spectrum No.~0 is used as the reference spectrum relative to which the differences of the wing-band mean fluxes of other spectra are calculated,
it does not appear in the plots in Figure \ref{fig:line_wings_consistency}b.
The consistency of the H$\alpha$ profiles in the line wings is judged by the maximum absolute difference value of the wing-band mean fluxes (indicated by a circle symbol in Figure \ref{fig:line_wings_consistency}b).

To determine the threshold of the maximum absolute difference of wing-band mean fluxe for recognizing the stellar sources with consistent H$\alpha$ profiles in the line wings,
we analyze the distribution of the maximum absolute difference values of the wing-band mean fluxes for the stellar source sample after the screening conditions described in Sections \ref{sec:sample_srccond_timedomain} and \ref{sec:sample_srccond_paraconsistency}.
The histograms of the distribution are shown in Figure \ref{fig:line_wings_consistency}c and \ref{fig:line_wings_consistency}d, 
which use linear and logarithmic scales on their vertical axes, respectively.
By carefully examining the distribution of the maximum absolute difference values of the wing-band mean fluxes and the associated spectral profiles,
the upper threshold of the maximum absolute difference of wing-band mean flux for recognizing the consistent H$\alpha$ profiles in the line wings is determined to be 0.022 (indicated with a horizontal dashed line in Figure \ref{fig:line_wings_consistency}b and a vertical dashed line in Figures \ref{fig:line_wings_consistency}c and \ref{fig:line_wings_consistency}d).
The H$\alpha$ spectra of a stellar source with the maximum absolute difference of wing-band mean flux greater than the threshold (such as the example shown in Figures \ref{fig:line_wings_consistency}a and \ref{fig:line_wings_consistency}b) are regarded to be inconsistent in the line wings.
Only the stellar sources with the maximum absolute difference of wing-band mean flux less than the threshold are kept in our sample.

\subsection{Consistency of H$\alpha$ profiles in the line center for the stellar sources with lower activity levels} \label{sec:appendix_profileconsistency_linecenter}

\begin{figure*}
  \centering
  \includegraphics[width=0.77\textwidth]{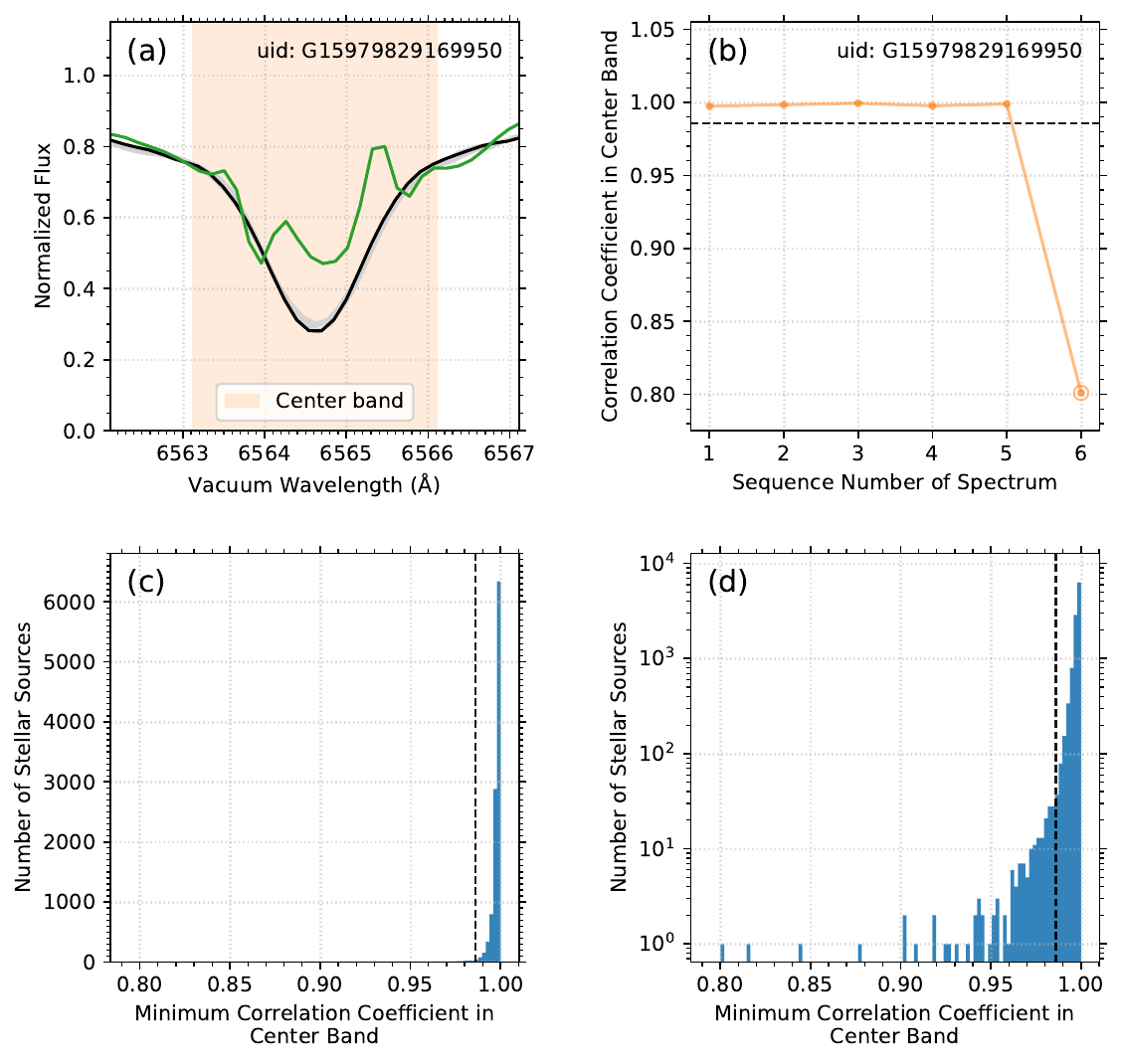}
  \caption{(a) Example MRS H$\alpha$ spectra of a lower activity level stellar source (uid: G15979829169950; seven spectra in total) that show inconsistent profiles in the line center.
  The spectral flux is normalized,
  and the wavelength shift caused by radial velocity is corrected.
  The spectra are sorted by their $I_\mathrm{H\alpha}$ index values.
  The spectrum with the smallest $I_\mathrm{H\alpha}$ index value (No. 0; bottom spectrum) is plotted in black,
  and the spectrum with the largest $I_\mathrm{H\alpha}$ index value (No. 6; top spectrum) is plotted in green.
  Other spectra (No. 1 to No. 5) are plotted in light gray.
  The 3\,{\AA}-wide center band of the H$\alpha$ line for testing the line-center consistency is indicated by a shaded area in orange.
  (b) Variation of the correlation coefficients of the normalized spectral fluxes in the center band with the sequence numbers of the spectra (sorted by $I_\mathrm{H\alpha}$ index).
  The spectrum No. 0 is used as the reference spectrum relative to which the correlation coefficients of other spectra are calculated,
  and hence does not appear in the plot.
  The circle symbol in the plot indicates the minimum correlation coefficient value. 
  (c) Distribution histogram of the minimum correlation coefficient values for the stellar source sample with lower activity levels after the screening conditions described in Sections \ref{sec:sample_srccond_timedomain} and \ref{sec:sample_srccond_paraconsistency},
  with a linear scale on the vertical axis.
  (d) Same as the histogram in panel (c), but with a logarithmic scale on the vertical axis.
  The horizontal dashed line in panel (b) and the vertical dashed lines in panels (c) and (d) indicate the lower threshold ($0.986$) of the minimum correlation coefficient in center band adopted in this work for recognizing the consistent H$\alpha$ profiles in the line center for the stellar sources with lower activity levels.
  } 
  \label{fig:line_center_consistency}
\end{figure*}

As described in Section \ref{sec:sample_srccond_profileconsistency}, 
in order to recognize the stellar objects with consistent H$\alpha$ profiles in the line center from the class of stellar sources with lower activity levels ($I_\mathrm{H\alpha}^\mathrm{median} < 0.5$),
we sort the H$\alpha$ spectra of multiple observations of each stellar source in the class by their $I_\mathrm{H\alpha}$ index values,
and calculate the correlation coefficients of the normalized spectral fluxes in a 3\,{\AA}-wide line-center band between the H$\alpha$ spectrum with the smallest $I_\mathrm{H\alpha}$ index value and other spectra with larger $I_\mathrm{H\alpha}$ index values.
Figure \ref{fig:line_center_consistency}a gives an example of the MRS H$\alpha$ spectra of a lower activity level stellar source (uid: G15979829169950; seven spectra in total) that show inconsistent profiles in the line center.
The spectrum with the smallest $I_\mathrm{H\alpha}$ index value (No. 0; bottom spectrum) is plotted in black, 
and the spectrum with the largest $I_\mathrm{H\alpha}$ index value (No. 6; top spectrum) is plotted in green.
Other spectra (No. 1 to No. 5) are plotted in light gray.
The center band of the H$\alpha$ line for calculating the correlation coefficients between the spectra is indicated by a shaded area in orange in Figure \ref{fig:line_center_consistency}a. 
The variation of the evaluated correlation coefficients with the sequence numbers of the spectra (sorted by $I_\mathrm{H\alpha}$ index) is shown in Figure \ref{fig:line_center_consistency}b.
Because the spectrum No. 0 is used as the reference spectrum relative to which the correlation coefficients of other spectra are calculated,
it does not appear in the plot in Figure \ref{fig:line_center_consistency}b.
The consistency of the H$\alpha$ profiles in the line center is judged by the minimum value of the evaluated correlation coefficients between the spectra (indicated by a circle symbol in Figure \ref{fig:line_center_consistency}b).

To determine the threshold of the minimum correlation coefficient for recognizing the lower activity level stellar sources with consistent H$\alpha$ profiles in the line center,
we analyze the distribution of the minimum correlation coefficient values for the stellar source sample with lower activity levels after the screening conditions described in Sections \ref{sec:sample_srccond_timedomain} and \ref{sec:sample_srccond_paraconsistency}.
The histograms of the distribution are shown in Figure \ref{fig:line_center_consistency}c and Figure \ref{fig:line_center_consistency}d,
which use linear and logarithmic scales on their vertical axes, respectively.
By carefully examining the distribution of the minimum correlation coefficient values and the associated spectral profiles,
the lower threshold of the minimum correlation coefficient for recognizing the consistent H$\alpha$ profiles in the line center for the stellar sources with lower activity levels is determined to be 0.986 
(indicated with a horizontal dashed line in Figure \ref{fig:line_center_consistency}b and a vertical dashed line in Figure \ref{fig:line_center_consistency}c and \ref{fig:line_center_consistency}d).
The H$\alpha$ spectra of a lower activity level stellar source with the minimum correlation coefficient less than the threshold (such as the example shown in Figures \ref{fig:line_center_consistency}a and \ref{fig:line_center_consistency}b) are regarded to be inconsistent in the line center.
Only the stellar objects in the lower activity level class with the minimum correlation coefficient in center band greater than the threshold are kept in our sample.

\section{$R_{\mathrm{H}{\alpha}}^\mathrm{STD} / R_{\mathrm{H}{\alpha}}^\mathrm{EXT}$ histograms of $N_\mathrm{\lowercase{obs}}=3$ and $N_\mathrm{\lowercase{obs}}=4$ of the whole MRS stellar source sample} \label{sec:appendix_histratiostdextrha_wholesample}

\begin{figure*}
  \centering
  \includegraphics[width=1.25\textwidth]{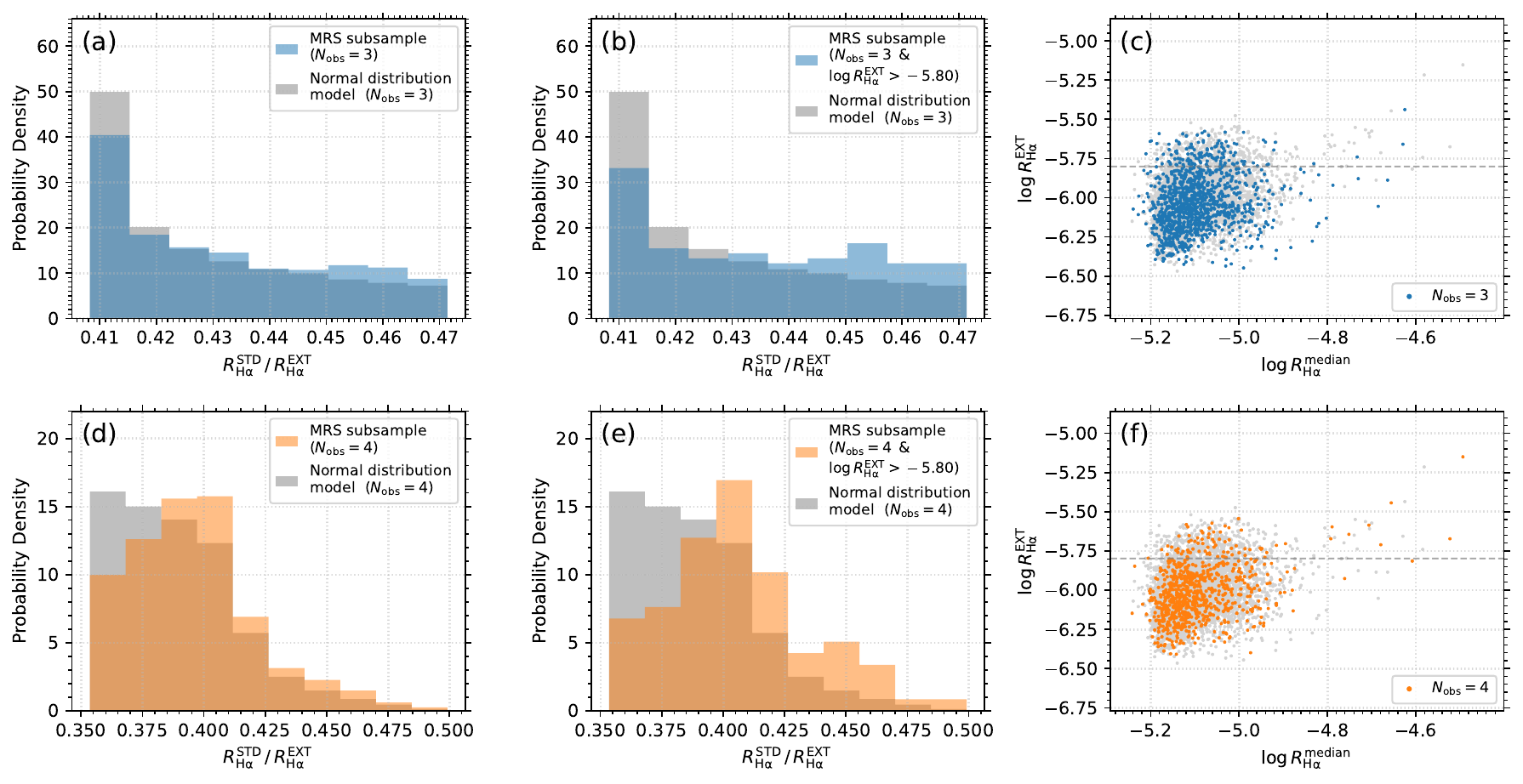}
  \caption{(leftmost column) Histograms of $R_{\mathrm{H}{\alpha}}^\mathrm{STD} / R_{\mathrm{H}{\alpha}}^\mathrm{EXT}$ for the MRS subsamples of stellar sources with $N_\mathrm{obs} = 3$ (top panel; in blue) and $N_\mathrm{obs} = 4$ (bottom panel; in orange) of the whole MRS stellar source sample, 
  compared with the simulated histograms of the normal distribution model (in gray). 
  (middle column) Same as in the leftmost column, but for the MRS subsamples after adding the screening condition of $\log R_\mathrm{H\alpha}^\mathrm{EXT} > -5.80$.
  (rightmost column) Scatter plots of $\log R_\mathrm{H\alpha}^\mathrm{EXT}$ versus $\log R_\mathrm{H\alpha}^\mathrm{median}$ for the MRS samples with $N_\mathrm{obs} = 3$ (top panel; in blue) and $N_\mathrm{obs} = 4$ (bottom panel; in orange) as employed in the leftmost column.
  The whole MRS stellar source sample is displayed in light gray in the background for reference.
  The horizontal dashed lines correspond to $\log R_\mathrm{H\alpha}^\mathrm{EXT} = -5.80$; 
  the data points above the lines are used in the middle column.
  }
  \label{fig:ratiostdextrha_histogram_wholesample}
\end{figure*}

In Section \ref{sec:discu_extrha_stdrha}, the $R_{\mathrm{H}{\alpha}}^\mathrm{STD} / R_{\mathrm{H}{\alpha}}^\mathrm{EXT}$ histograms of $N_\mathrm{obs}=3$ and $N_\mathrm{obs}=4$ are displayed for the MRS stellar sources with $\log R_\mathrm{H\alpha}^\mathrm{median} < -4.85$ (see Figure \ref{fig:ratiostdextrha_histogram}) to avoid the disturbance of the $\log R_\mathrm{H\alpha}^\mathrm{median} > -4.85$ stellar sources to the analysis result. 
As a comparison, in Figure \ref{fig:ratiostdextrha_histogram_wholesample} we show the $R_{\mathrm{H}{\alpha}}^\mathrm{STD} / R_{\mathrm{H}{\alpha}}^\mathrm{EXT}$ histograms of $N_\mathrm{obs}=3$ and $N_\mathrm{obs}=4$ of the whole MRS stellar source sample ($\log R_\mathrm{H\alpha}^\mathrm{median} < -4.85$ stellar sources plus $\log R_\mathrm{H\alpha}^\mathrm{median} > -4.85$ stellar sources).    

The layout of Figure \ref{fig:ratiostdextrha_histogram_wholesample} is similar to that of Figure \ref{fig:ratiostdextrha_histogram}.
The leftmost column of Figure \ref{fig:ratiostdextrha_histogram_wholesample} shows the histograms of $R_{\mathrm{H}{\alpha}}^\mathrm{STD} / R_{\mathrm{H}{\alpha}}^\mathrm{EXT}$ for the MRS subsamples of stellar sources with $N_\mathrm{obs}=3$ (top panel; in blue) and $N_\mathrm{obs}=4$ (bottom panel; in orange) of the whole MRS stellar source sample, 
compared with the simulated histograms of the normal distribution model (in gray).
The middle panel of Figure \ref{fig:ratiostdextrha_histogram_wholesample} shows the histograms of $R_{\mathrm{H}{\alpha}}^\mathrm{STD} / R_{\mathrm{H}{\alpha}}^\mathrm{EXT}$ for the MRS subsamples of $N_\mathrm{\lowercase{obs}}=3$ and $N_\mathrm{\lowercase{obs}}=4$ of the whole sample after adding the screening condition of $\log R_\mathrm{H\alpha}^\mathrm{EXT} > -5.80$.
The rightmost column of Figure \ref{fig:ratiostdextrha_histogram_wholesample} shows the scatter plots of $\log R_\mathrm{H\alpha}^\mathrm{EXT}$ versus $\log R_\mathrm{H\alpha}^\mathrm{median}$ for the MRS subsamples of $N_\mathrm{\lowercase{obs}}=3$ and $N_\mathrm{\lowercase{obs}}=4$ used in the histograms. 

By comparing the histograms of $R_{\mathrm{H}{\alpha}}^\mathrm{STD} / R_{\mathrm{H}{\alpha}}^\mathrm{EXT}$ in Figures \ref{fig:ratiostdextrha_histogram} and \ref{fig:ratiostdextrha_histogram_wholesample},
it can be seen that after adding the $\log R_\mathrm{H\alpha}^\mathrm{median} > -4.85$ stellar sources to the $\log R_\mathrm{H\alpha}^\mathrm{median} < -4.85$ stellar sources,
the changes of the histograms are subtle in panels (a), (b), and (d) of Figure \ref{fig:ratiostdextrha_histogram_wholesample},
which is owing to the small sample size of the $\log R_\mathrm{H\alpha}^\mathrm{median} > -4.85$ stellar sources relative to the $\log R_\mathrm{H\alpha}^\mathrm{median} < -4.85$ stellar sources (see the scatter plots in Figures \ref{fig:ratiostdextrha_histogram_wholesample}c and \ref{fig:ratiostdextrha_histogram_wholesample}f). 
The prominent change of the $R_{\mathrm{H}{\alpha}}^\mathrm{STD} / R_{\mathrm{H}{\alpha}}^\mathrm{EXT}$ distribution appears in panel (e) of Figure \ref{fig:ratiostdextrha_histogram_wholesample}, 
which corresponds to the MRS subsample with $N_\mathrm{obs}=4$ \& $\log R_\mathrm{H\alpha}^\mathrm{EXT} > -5.80$.
The reason is that for this subsample of stellar sources, the sample size of the $\log R_\mathrm{H\alpha}^\mathrm{median} > -4.85$ stellar sources is relatively large (see Figure \ref{fig:ratiostdextrha_histogram_wholesample}f).

As exhibited in Figure \ref{fig:ratiostdextrha_histogram_wholesample}e,
after adding the $\log R_\mathrm{H\alpha}^\mathrm{median} > -4.85$ stellar sources,
the discrepancies between the $R_{\mathrm{H}{\alpha}}^\mathrm{STD} / R_{\mathrm{H}{\alpha}}^\mathrm{EXT}$ histograms of the MRS sample and the normal distribution model are reduced.
This result suggests that the H$\alpha$ variations of the $\log R_\mathrm{H\alpha}^\mathrm{median} > -4.85$ stellar sources are more likely to be random fluctuations.

\section{Dataset of this work} \label{sec:appendix_dataset}

The dataset of this work includes a catalog of the measures of H$\alpha$ chromospheric activity intensity and H$\alpha$ variability for the stellar sources in the MRS time-domain sample of solar-like stars employed in this work.
The data of the MRS spectra of multiple observations for each stellar source are also included in the catalog.
The catalog is stored in CSV format,
and the file name of the catalog is {\tt\string Halpha\_variability\_catalog\_LAMOST\_MRS\_DR10.csv}. 
The columns contained in the catalog are tabulated in Table \ref{tab:catalogsources}.
The dataset is available online at \url{https://doi.org/10.5281/zenodo.17083328}.

\begin{table}[tpb]
\centering
\caption{Columns contained in the catalog of the dataset.}
\label{tab:catalogsources}
\begin{tabular*}{\textwidth}{@{\extracolsep\fill}lcp{7.70cm}}
  \toprule
   Column & Unit & Description   \\
  \hline
   {\tt\string uid} &  & unique LAMOST source identifier  \\
   {\tt\string ra} & degree & right ascension of each stellar source  \\
   {\tt\string dec} & degree & declination of each stellar source  \\
   {\tt\string teff} & K & effective temperature ($T_\mathrm{eff}$) of each stellar source \\
   {\tt\string teff\_err} & K & uncertainty of $T_\mathrm{eff}$ \\
   {\tt\string logg} & dex & surface gravity ($\log\,g$) of each stellar source \\
   {\tt\string logg\_err} & dex & uncertainty of $\log\,g$ \\
   {\tt\string feh} & dex & metallicity ($\mathrm{[Fe/H]}$) of each stellar source \\
   {\tt\string feh\_err} & dex & uncertainty of $\mathrm{[Fe/H]}$ \\
   {\tt\string N\_obs} &  & number of MRS observations ($N_\mathrm{obs}$) of each stellar source \\
   {\tt\string T\_span} & day & time span of multiple observations ($T_\mathrm{span}$) of each stellar source \\ 
   {\tt\string obsid\_list} &  & list of {\tt\string obsid} (unique LAMOST observation identifier) for multiple observations of each stellar source \\
   {\tt\string obstime\_list} &  & list of observation time (UTC) for multiple observations of each stellar source \\
   {\tt\string mjd\_list} & day & list of Modified Julian Day (MJD) for multiple observations of each stellar source \\
   {\tt\string I\_halpha\_list} &  & list of H$\alpha$ activity index ($I_{\mathrm{H}{\alpha}}$) for multiple observations of each stellar source \\
   {\tt\string I\_halpha\_err\_list} &  & list of uncertainty of $I_{\mathrm{H}{\alpha}}$ ($\delta I_{\mathrm{H}{\alpha}}$) for multiple observations of each stellar source \\
   {\tt\string median\_I\_halpha} &  & median $I_{\mathrm{H}{\alpha}}$ ($I_{\mathrm{H}{\alpha}}^\mathrm{median}$) for multiple observations of each stellar source \\
   {\tt\string median\_I\_halpha\_err} &  & uncertainty of $I_{\mathrm{H}{\alpha}}^\mathrm{median}$ ($\delta I_{\mathrm{H}{\alpha}}^\mathrm{median}$) \\ 
   {\tt\string R\_halpha\_list} &  & list of H$\alpha$ $R$-index ($R_{\mathrm{H}{\alpha}}$) for multiple observations of each stellar source \\
   {\tt\string R\_halpha\_err\_list} &  & list of uncertainty of $R_{\mathrm{H}{\alpha}}$ ($\delta R_{\mathrm{H}{\alpha}}$) for multiple observations of each stellar source  \\
   {\tt\string median\_R\_halpha} &  & median $R_{\mathrm{H}{\alpha}}$ ($R_{\mathrm{H}{\alpha}}^\mathrm{median}$) for multiple observations of each stellar source \\
   {\tt\string median\_R\_halpha\_err} &  & uncertainty of $R_{\mathrm{H}{\alpha}}^\mathrm{median}$ ($\delta R_{\mathrm{H}{\alpha}}^\mathrm{median}$) \\ 
   {\tt\string EXT\_R\_halpha} &  & extent of $R_{\mathrm{H}{\alpha}}$ fluctuation ($R_{\mathrm{H}{\alpha}}^\mathrm{EXT}$) for multiple observations of each stellar source \\
   {\tt\string EXT\_R\_halpha\_err} &  & uncertainty of $R_{\mathrm{H}{\alpha}}^\mathrm{EXT}$ ($\delta R_{\mathrm{H}{\alpha}}^\mathrm{EXT}$) \\
   {\tt\string EXT\_R\_halpha\_gt\_3err} &  & indicating if the value of $R_{\mathrm{H}{\alpha}}^\mathrm{EXT}$ is greater than $3 \times \delta R_{\mathrm{H}{\alpha}}^\mathrm{EXT}$; `{\tt\string Y}' representing yes and `{\tt\string N}' representing no \\
   {\tt\string STD\_R\_halpha} &  & STD (standard deviation) of $R_{\mathrm{H}{\alpha}}$ fluctuation  ($R_{\mathrm{H}{\alpha}}^\mathrm{STD}$) for multiple observations of each stellar source \\
  \botrule
\end{tabular*}
\footnotetext{
  Note: 
  (1) The observational and spectroscopic parameters are from the MRS catalogs and FITS files of LAMOST DR10 v1.0; 
  the parameter values of each stellar source are the medians of multiple observations. 
  (2) The data in a list of parameter values for multiple observations of each stellar source are sorted by observation time.}
\end{table}

\end{appendices}

\backmatter

\newpage

\bmhead{Acknowledgements}
Guoshoujing Telescope (the Large Sky Area Multi-Object Fiber Spectroscopic Telescope, LAMOST) is a National Major Scientific Project built by the Chinese Academy of Sciences. Funding for the project has been provided by the National Development and Reform Commission. LAMOST is operated and managed by the National Astronomical Observatories, Chinese Academy of Sciences.
This work made use of Astropy \citep{2013A&A...558A..33A,2018AJ....156..123A} and SciPy \citep{2020NatMe..17..261V}.

\bmhead{Author contributions}
Han He performed the data analysis and wrote the manuscript with input from all coauthors.

\bmhead{Funding}
This research was supported by the National Key R\&D Program of China (2019YFA0405000) and the National Natural Science Foundation of China (11973059).

\bmhead{Data Availability}
The dataset generated during the current study is available online (see Appendix \ref{sec:appendix_dataset} for the web link).

\vspace{5mm}

\section*{Declarations}

\bmhead{Informed Consent}
Informed consent was obtained from all individual participants included in the study.

\bmhead{Competing interests}
The authors declare no competing interests.

\end{document}